\documentclass[aps,prb,onecolumn,nofootinbib,citeautoscript,10pt]{revtex4-2}

\usepackage{amsmath,amssymb,bm} 
\usepackage{graphicx,comment}

\usepackage[tight]{subfigure} 

\usepackage[dvipsnames]{xcolor} 
\usepackage[papersize={8.5in,11in}]{geometry}
\usepackage[colorlinks=true]{hyperref}
\hypersetup{
    bookmarks=true,         
    unicode=false,          
    pdftoolbar=true,        
    pdfmenubar=true,        
    pdffitwindow=false,     
    pdfstartview={FitH},    
    pdfkeywords={keyword1} {key2} {key3}, 
    pdfnewwindow=true,      
    colorlinks=true,       
    linkcolor=magenta, 
    citecolor=blue,        
    filecolor=magenta,      
    urlcolor=blue           
} 

\usepackage{dcolumn}
\usepackage{color}
\usepackage{amssymb,amsmath, makecell}
\usepackage{tabularx, makecell}
\usepackage{epstopdf}
\usepackage{latexsym}
\usepackage{colortbl}
\usepackage{psfrag}
\usepackage{bbm,bm,array,physics}
\usepackage{dsfont}
\usepackage{float, mathrsfs}

\def \nn{\nonumber \\}

\DeclareRobustCommand{\bs}[1]{\bm{#1}}

\begin{document}
\title{Interplay of strain-induced axial gauge fields and intrinsic band-topology in the magnetoelectric conductivity of gapped nodal rings}

\author{Firdous Haidar}
\author{Muhammed Jaffar A.}
\author{Ipsita Mandal}
\email{ipsita.mandal@snu.edu.in}

\affiliation{Department of Physics, Shiv Nadar Institution of Eminence (SNIoE), Gautam Buddha Nagar, Uttar Pradesh 201314, India}

\begin{abstract}
We compute the magnetoelectric conductivity of a semimetal hosting an ideal gapped nodal ring (GNR) in three distinct planar-Hall configurations, in the simultaneous presence of an external electric field $\boldsymbol{E}$, a magnetic field $\boldsymbol{B}$, and a
strain-induced axial pseudomagnetic field $\boldsymbol{B}_5$. The latter arises from a nonuniform lattice deformation and couples to antipodal points on the toroidal Fermi surface with opposite signs, reflecting its chiral nature. Extending our earlier analysis to include $\boldsymbol{B}_5$, we demonstrate how its vortex-like field lines --- co-aligned with the Berry curvature (BC) and orbital magnetic moment (OMM) --- imprint qualitatively distinct signatures on the conductivity tensor. In particular, this alignment causes the dot product of $\boldsymbol{B}_5$ with the BC or OMM-induced quantities to be angle-independent on the Fermi surface, generating a nonvanishing integral linear-in-$B_5$, which is not possible for isotropic nodal points harbouring BC-monopoles. We show that a part of the planar-Hall conductivity in the first set-up remains completely immune to strain, providing a strain-insensitive internal reference for topological transport. Our explicit analytical expressions offer concrete and experimentally-testable predictions for identifying strain-induced signatures in transport measurements on GNR materials.
\end{abstract}

\maketitle

\tableofcontents

\section{Introduction}

\begin{figure*}[t]
\subfigure[]{\includegraphics[width=0.25 \textwidth]{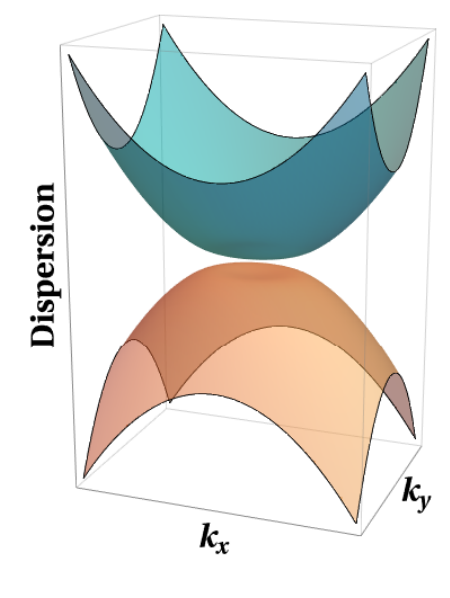}} \hspace{ 1 cm}
\subfigure[]{\includegraphics[width=0.3 \textwidth]{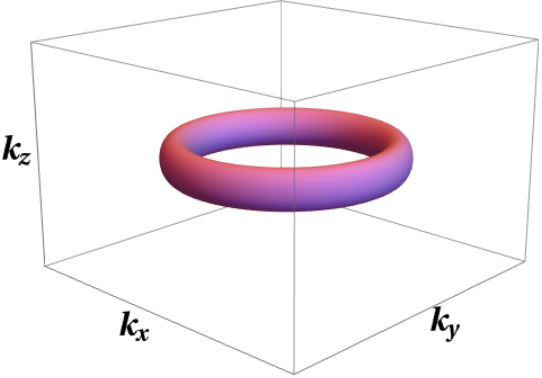}  
\includegraphics[width=0.3 \textwidth]{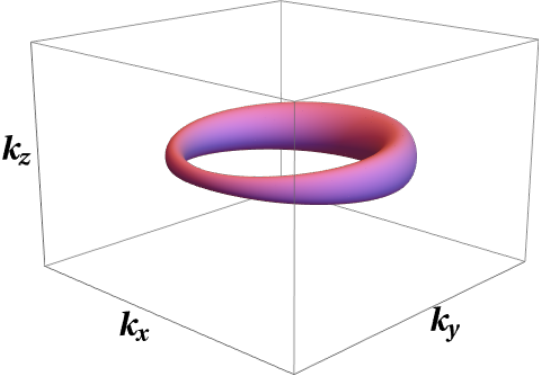}}
\caption{\label{figfs}A gapped nodal-ring semimetal with isotropy in the plane of the
nodal-loop defined by $k_z = 0$: (a) The band dispersion as a function of momentum in 
the $k_x k_y$-plane. (b) A schematic comparison of the Fermi surface 
topology in the two cases of vanishing and nonvanishing OMM correction. 
The external magnetic field $\bs{B}$ is directed along the $y$-axis 
throughout. The original toroidal Fermi surface is distorted into a 
ring cyclide upon switching on $\bs{B}$, and we restrict to field 
strengths, $|\bs{B}|$, small-enough such that the system remains below the 
threshold for a Lifshitz transition to a horn-cyclide geometry.}
\end{figure*}

 The discovery of three-dimensional (3d) materials called semimetals, which possess symmetry-protected band-crossings in their 3d Brillouin zones (BZs), has opened a direct pathway from the abstract mathematics of topology to the concrete physics of material bandstructures \cite{burkov11_Weyl, yan17_topological, bernevig, armitage_review, ips-hermann-review}. Those remarkable systems have exemplified the materials whose BZs, viewed as closed manifolds, are endowed with the nontrivial topological character \cite{zak_nodal, fu_nlsm, chiu_classification}. Depending on the geometry of the band-crossings, they are found to occur at isolated nodal points \cite{burkov11_Weyl, yan17_topological, bernevig} or along the extended nodal curves \cite{balents-nodal}, giving rise to the zero-dimensional or one-dimensional (1d) Fermi surfaces, respectively, when the chemical potential is tuned to the band-crossing energy. In either case, these band-crossings mark singular points of the BZ --- 
spanned by the momentum coordinates $\bs{k} \equiv \{k_x, k_y, k_z\}$ 
--- where the density of states vanishes \cite{schnyder_nodal, fu_nlsm}. Nodal points behave as 
sources and sinks of the Berry curvature (BC), whereas nodal rings 
(NRs) carry a vanishing Chern number, their topological character 
instead being encoded in a quantized Zak phase \cite{schnyder_nodal, 
zak_nodal, fu_nlsm} equal to an integer multiple of $\pi$ 
\cite{schnyder_nodal, biao_nodal, zak_nodal}. An NR itself 
can be visualised as the locus at which the BC diverges, while it vanishes identically away from it \cite{ips-nlsm-ph, ips-dipole-vnr}. Introducing a $\mathcal{P}\mathcal{T}$-symmetry-breaking mass term 
($\propto \Delta$) gaps out the NR, after which the BC becomes 
nonvanishing and smooth throughout the entire BZ \cite{ips-nlsm-ph, ips-dipole-vnr}. A finite $\Delta$ 
thereby transforms the 1d nodal-ring Fermi surface into a 
two-dimensional (2d) toroidal manifold that wraps around the original nodal 
circle \cite{yang1, ips-nlsm-ph}. Throughout this paper, we take the nodal circle to lie in a 
plane perpendicular to $k_z$, with full rotational symmetry about the 
$k_z$-axis (cf.\ Fig.~\ref{figfs}). The resulting gapped nodal ring 
(GNR) supports BC flux lines that wind into vortices around the 
$k_z$-axis \cite{ips-nlsm-ph, ips-dipole-vnr}.

At the heart of all topological phenomena in the semimetals is the Berry phase, which 
gives rise to the BC and other geometrical quantities defined over the 
BZ \cite{xiao_review, sundaram99_wavepacket, timm, ips_rahul_ph_strain, 
graf-Nband, rahul-jpcm, ips-kush-review, ips-ruiz, 
ips-rsw-ph, ips-tilted, ips-spin1-ph}. When a semimetal is 
placed in a nonzero magnetic field, the Berry phase also generates the 
orbital magnetic moment (OMM), which arises from the semiclassical 
self-rotation of the quasiparticle wavepacket \cite{xiao_review, 
sundaram99_wavepacket} and leaves its mark on a wide range of 
observables. Prominent examples of linear transport phenomena, shaped by the 
BC and OMM, include the intrinsic anomalous-Hall effect 
\cite{haldane04_berry, goswami13_axionic, burkov14_anomolous} and 
planar-Hall conductivity \cite{zhang16_linear, chen16_thermoelectric, 
nandy_2017_chiral, amit_magneto, pal22a_berry, fu22_thermoelectric, 
 mizuta14_contribution, ips-mwsm-floquet, 
ips_rahul_ph_strain, timm, rahul-jpcm, ips-kush-review, 
claudia-multifold, ips-ruiz, phe_nlsm, ips-tilted, ips-rsw-ph, ips-spin1-ph}. The topological properties of 3d nodal-ring semimetals manifest in all 
these transport channels and GNRs, in particular, give rise to 
distinctive features in the Berry-phase-induced response 
coefficients \cite{schnyder_nodal, yang1, yang_review_nlsm, chen_nlsm, 
ips-magnus, phe_nlsm, claudia_nlsm, enke, ips-nlsm-ph, ips-dipole-vnr}. A notable 
consequence is that, by contributing significant BC over an extended 
region of the BZ, a GNR can substantially amplify the anomalous-Hall 
response \cite{enke}.

Experimental realizations of GNRs span a broad range of materials, 
including SrAs$_3$ \cite{arpes-nlsm}, Ca$_3$P$_2$ \cite{expt1_nlsm}, 
hexagonal pnictides such as CaAgP and CaAgAs \cite{expt2_nlsm}, 
photonic metamaterials \cite{biao_nodal}, alkaline-earth metals 
including Ca, Sr, and Yb \cite{alkaline_nlsm}, Fe$_2$MnX 
\cite{claudia_nlsm}, and Co$_3$Sn$_2$S$_2$ \cite{enke}. First-principles 
calculations predict that CuTeO$_3$ \cite{cuteo_nlsm} hosts an ideal 
GNR: the nodal loop lies close to the Fermi level, is nearly flat in 
energy and confined to the $k_x k_y$-plane, well approximated as 
circular, and unperturbed by extraneous bands. Spin-orbit coupling is negligible in this compound, making it an excellent embodiment of the model Hamiltonian we consider here and 
lending concrete justification to our idealisation of a circular nodal 
curve as depicted in Fig.~\ref{figfs}. Last but not the least, GNRs can be realised in 
an acoustic crystal \cite{nlr-acoustic}.

Lattice strain in Dirac and Weyl semimetals can be recast as an effective chirality-dependent elastic gauge field~\cite{landsteiner_gauge, liu_gauge}, giving rise to a chiral anomaly under strain-induced pseudomagnetic and pseudoelectric fields, $\bs{B}_5$ and $\bs{E}_5$~\cite{pikulin_gauge}. A uniform pseudomagnetic field 
$\bs{B_5}$ can be generated by twisting a Weyl or multi-Weyl semimetal nanowire under torsion, while a pseudoelectric field $\bs{E_5} $ arises  from oscillatory compression and extension of the crystal along an 
axis, as can be achieved by driving longitudinal sound waves \cite{pikulin_gauge}. In other words, torsional and time-dependent lattice deformations offer a route to generating such pseudogauge fields~\cite{morales23_curvature}. Strain-induced pseudomagnetic fields have since been generated experimentally in a doped type-II Weyl semimetal~\cite{exp_gauge}, and pseudo-Landau levels arising from an analogous gauge field have been realised in a 3d inhomogeneous acoustic crystal~\cite{nlr-acoustic}. On the transport side, strain has been shown to modify the electric, thermal, and thermoelectric planar-Hall response of Weyl and multi-Weyl semimetals~\cite{ips_rahul_ph_strain, ips-ruiz, ips-rsw-ph}. This construction extends to nodal-line semimetals, where a time-dependent strain gives rise to a pseudoelectric field, which drives a topological piezoelectric response tied to a parity anomaly along the nodal line~\cite{nlsm-strain}.
Pseudoelectromagnetic fields can thus be induced in GNRs as well through the same lattice-deformation route \cite{nlsm-strain, pikulin-gauge-nlsm} --- the coupling between elastic strain and low-energy electronic excitations then generate effective pseudogauge fields \cite{guinea10_energy, guinea10_generating, low10_strain, landsteiner_gauge, liu_gauge, 
pikulin_gauge, onofre, ips_rahul_ph_strain, ips-ruiz, nlsm-strain}. A defining 
feature of these emergent axial fields is that they couple to 
excitations at antipodal points of the toroidal Fermi surface with opposite signs, 
in sharp contrast with physical electromagnetic fields, which couple 
universally with the same sign \cite{pikulin_gauge, nlsm-strain, ips-dipole-vnr}. Experimental confirmation of pseudoelectromagnetic-field generation in doped semimetals and acoustic crystals has been reported in Refs.~\cite{exp_gauge, nlr-acoustic}.

The focus of this paper is the analytical computation of the 
linear response in the form of magnetoelectric conductivity of an ideal GNR, when subjected 
simultaneously to an external electric field $\bs{E}$, a magnetic field 
$\bs{B}$, and a nonuniform mechanical strain that induces an effective gauge field, $\bs{A_5} $ [cf. Fig.~\ref{figsetup}(a)]. $\bs{A_5} $ gives rise to an effective axial magnetic field, $\bs{B_5}$. 
While $\bs{B_5} = \nabla_{\bs r} \cross \boldsymbol{ A_5} $ is confined to the parent NR's plane,
with its field lines forming vortices \cite{ips-dipole-vnr, nlsm-strain},
we vary the orientation of the $\bs E  \bs B$-plane
and investigate three distinct planar-Hall set-ups as shown in Fig.~\ref{figsetup}(b).
The unstrained response for all these set-ups was already investigated in our earlier work \cite{ips-nlsm-ph}. In this paper, the central question that we address is what kind of detectable imprint $\bs{B_5}$
leaves on the resulting linear response for each individual set-up.
This is motivated by the crucial feature that, 
while $\bs B$ couples with all points on the GNR's Fermi surface with the same sign, 
$\bs B_5$ does so in a chiral way (i.e., the couplings with 
antipodal regions appear with opposite signs) \cite{ips-dipole-vnr, nlsm-strain}.
Our work is also motivated by the known
enhancement of the planar-Hall conductivity by terms $\propto \bs{B_5}^2$  in nodal-point 
semimetals \cite{grushin-strain-wsm, rahul-jpcm, ips-ruiz, ips-rsw-ph}.

To place the present work in context, we note that earlier studies of transport have addressed the anomalous-Hall response of GNRs and related nodal-line materials~\cite{enke, claudia_nlsm}, the thermopower of nodal-line semimetals~\cite{chakraborty_24}, the planar-Hall effect for semimetals and GNRs in the presence of physical electric and magnetic fields alone~\cite{ips-tilted, ips-spin1-ph, ips-nlsm-ph, phe_nlsm}, and the planar-Hall effect for nodal-point semimetals under strain-induced axial gauge fields~\cite{onofre, ips_rahul_ph_strain, ips-ruiz, ips-rsw-ph}. Furthermore, 3d pseudo-Landau levels generated by an inhomogeneous lattice-distortion-induced pseudomagnetic field have been realised experimentally in an acoustic crystal~\cite{nlr-acoustic}. In Ref.~\cite{ips-nlsm-ph}, the dependence of the planar-Hall response for an ideal GNR on the direction of the applied fields has been computed. Beyond these field-dependent linear-response studies, the intrinsic transport properties of GNRs have also been investigated theoretically, including the combined a.c. and d.c. conductivity at the node~\cite{mukherjee17_transport_optics}, the anomalous Coulomb screening and disorder-limited conductivity of a nodal line~\cite{syzranov17_transport}, the disorder-driven multifractality of the nodal-loop wavefunction~\cite{goncalves20_multifractality}, and the intrinsic and extrinsic contributions to the longitudinal d.c. conductivity of a mass-gapped Dirac nodal-line semimetal~\cite{pandey24_longitudinal_dc}. However, none of these treats the magnetoelectric response of a GNR in the simultaneous presence of a physical magnetic field $\boldsymbol{B}$ and a strain-induced axial pseudomagnetic field $\boldsymbol{B}_5$. Nor did anyone identify the angle-independent alignment of $\boldsymbol{B}_5$ with the BC and OMM on the toroidal Fermi surface, that we show below gives rise to a term linear-in-$B_5$ -- a feature with no counterpart for isotropic Weyl nodes. To the best of our knowledge, this interplay, together with the resulting strain-insensitive channel identified in set-up~I, has not been reported previously, and constitutes the central new result of this paper.

The paper is organised as follows. In Sec.~\ref{secmodel}, we introduce the model Hamiltonian for a GNR and define the structure of the pseudomagnetic field. In Sec.~\ref{secsigma}, we elaborate on our formalism as well as compute the explicit analytical expressions of all the conductivity components in each set-up. Section~\ref{secdis} is devoted to a thorough analysis of the results obtained in Sec.~\ref{secsigma}.
We end with a summary and outlook in Sec.~\ref{secsum}. Some long expressions and intermediate steps are demonstrated in the appendices. Throughout what follows, we work in natural units, setting the reduced 
Planck constant $\hbar$, the speed of light $c$, and the Boltzmann 
constant $k_B$ each equal to unity. In these units, the magnitude of 
the electric charge $e$ is likewise dimensionless and equal to unity; 
nevertheless, we retain $e$ explicitly in all expressions for the 
purpose of book-keeping.

\section{Model}
\label{secmodel}

The minimal two-band model of a GNR with a single circular nodal loop 
lying in the $k_x k_y$-plane is given by \cite{balents-nodal, yang1}
\begin{align}
\mathcal{H}_0(\bs{k}) = \bs{d}_0(\bs{k})\cdot\boldsymbol{\sigma}\,, \quad
\bs{d}_0(\bs{k}) = \left\{\lambda\left(k_\perp^2 - k_0^2\right),\,
v_z\,k_z,\,\Delta\right\}, \quad
k_\perp = \sqrt{k_x^2 + k_y^2}\,,
\end{align}
where $\boldsymbol{\sigma} = \{\sigma_x, \sigma_y, \sigma_z\}$ denotes 
the vector of Pauli matrices. The parameters $\lambda$ and $k_0$ are 
material-dependent, while $\Delta$ encodes the small gap induced by 
symmetry breaking, for instance through inversion-breaking uniaxial
strain, pressure, or an external electric field \cite{schnyder_nodal}. In the 
ungapped limit $\Delta = 0$, the two bands cross along the locus 
$k_\perp^2 - k_0^2 = 0$, which defines a nodal ring of radius $k_0$. 
When the chemical potential satisfies $\mu \ll \lambda\,k_0^2$, the 
relevant low-energy excitations are confined to the neighborhood of 
the Fermi surface encircling the nodal ring. To characterise the 
transport signatures of these quasiparticles, it is convenient to 
linearise $\mathcal{H}$ in the momentum deviation from the nodal curve 
\cite{linearize-nlsm}, which is most naturally accomplished by 
introducing toroidal coordinates:
\begin{align}
\label{eqtrs}
k_x = \left(k_0 + \kappa\cos\gamma\right)\cos\phi\,, \quad
k_y = \left(k_0 + \kappa\cos\gamma\right)\sin\phi\,, \quad
k_z = \frac{\kappa\sin\gamma}{\alpha}\,, \quad
\alpha = \frac{v_z}{v_0}\,, \quad v_0 = 2\lambda\,k_0\,.
\end{align}
The Jacobian of this coordinate transformation is 
$J = \kappa\left(k_0 + \kappa\cos\gamma\right)/\alpha$. Inverting the 
transformation gives $k_0 + \kappa\cos\gamma = \pm\,k_\perp$, and 
since $\kappa \ll k_0$ in the low-energy limit, we have
$ \kappa \cos  \gamma   = k_\perp - k_0 $.
This leads to
\begin{align}
& \mathcal{H}_0 (\bs k ) = 
\mathcal{H} (\delta \bs k) + \order{\kappa^2}\,, \quad 
\mathcal{H} (\delta \bs k) = 
{\bs d} ( \delta \bs k) \cdot \boldsymbol{\sigma} \,,
\quad
\delta \bs k =
\kappa  \left \lbrace \cos  \gamma  \cos  \phi , \,
\cos  \gamma  \sin  \phi , \,
\frac{\sin  \gamma } {\alpha} 
 \right \rbrace, \nn 
& \bs{d}( \delta \bs{k}) 
= \left \lbrace v_0 \, \kappa \cos  \gamma  ,
\, v_0 \, \kappa \sin  \gamma ,\, \Delta \right\rbrace
=
\left \lbrace v_0  \left( k_\perp - k_0 \right) ,
\, v_z \, k_z ,\, \Delta \right\rbrace .
\end{align}
In the toroidal coordinate system, $k_0$ represents the major radius 
of the torus --- the distance from the centre of the tube to the centre 
of the nodal ring --- while $\kappa$ plays the role of the minor radius, 
measuring the cross-sectional extent of the torus. The toroidal angle 
$\phi$ and the poloidal angle $\gamma$, each ranging over $[0, 2\pi)$, 
describe rotations around the nodal ring and around the torus's axis 
of revolution, respectively. The parameter $\alpha$ captures the 
anisotropy of the system, encoding the ratio of the velocity along the 
$k_z$-axis to that within the $k_x k_y$-plane.

Focusing on the linearised Hamiltonian $\mathcal{H}$, the eigenvalues of the two bands are found to be
\begin{align}
\label{6}
\varepsilon_s  ({ \bs k}) &=(-1)^s\, \epsilon, \quad 
 \epsilon = \sqrt{ v_0^2 \,\kappa^2 +  \Delta^2}, 
\quad
s \in \lbrace 1,2 \rbrace,
\end{align}
where $s = 1$ and $s = 2$ label the valence (negative-energy) and 
conduction (positive-energy) bands, respectively. The band-velocity 
of the quasiparticles takes the form of
\begin{align}
\label{eqv}
{\boldsymbol v}^{(0,s)} ( \bs{k}) 
\equiv \nabla_{\bs{k}} \varepsilon_s  (\bs{k}) = 
\frac{(-1)^s \, v_0^2}{\epsilon} 
\left \lbrace k_x \left( 1- \frac{k_0}{k_\perp } \right ), 
\, k_y \left( 1- \frac{k_0} {k_\perp} \right), 
\, \frac{ v_z^2 \, k_z} {v_0^2 }  \right\rbrace.
\end{align}

\subsection{Topological quantities affecting transport}

The Berry curvature (BC) and the orbital magnetic moment (OMM), associated 
with the $s^{\text{th}}$ band, are evaluated using the general expressions,
\begin{align}
\label{eqomm}
& \bs{\Omega}_s(\bs{k}) = i\,\langle\nabla_{\bs{k}}\psi_s(\bs{k})|\,
\cross\,|\nabla_{\bs{k}}\psi_s(\bs{k})\rangle
\Rightarrow
\Omega^i_s(\bs{k})
\overset{\text{two}-}{\underset{\text{band}}{=}}
\frac{(-1)^{s+1}\,\epsilon^i_{\;\;jl}}
{4\,|\bs{d}(\delta\bs{k})|^3}\,\bs{d}(\delta\bs{k})\cdot
\left[\partial_{k_j}\bs{d}(\delta\bs{k})\cross
\partial_{k_l}\bs{d}(\delta\bs{k})\right]
\quad \text{and}
\nonumber \\
& \boldsymbol{m}_s(\bs{k}) = \frac{-i\,e}{2}\,
\langle\bs{\nabla}_{\bs{k}}\psi_s(\bs{k})|\cross
\left[\left\{\mathcal{H}(\bs{k}) - \mathcal{E}_s(\bs{k})\right\}
|\bs{\nabla}_{\bs{k}}\psi_s(\bs{k})\rangle\right]
\Rightarrow
m^i_s(\bs{k})
\overset{\text{two}-}{\underset{\text{band}}{=}}
\frac{-e\,\epsilon^i_{\;\;jl}}
{4\,|\bs{d}(\delta\bs{k})|^2}\,\bs{d}(\delta\bs{k})\cdot
\left[\partial_{k_j}\bs{d}(\delta\bs{k})\cross
\partial_{k_l}\bs{d}(\delta\bs{k})\right],
\end{align}
respectively. Here $|\psi_s(\bs{k})\rangle$ denotes the normalised 
eigenvector of the band labelled by $s$, with $\{|\psi_1\rangle,\,
|\psi_2\rangle\}$ forming an orthonormal set. For two-band models of 
the generic form $\bs{d}\cdot\boldsymbol{\sigma}$, one has the 
relation $m^i_s(\bs{k}) = e\,\varepsilon_s(\bs{k})\,\Omega^i_s(\bs{k})$ 
\cite{graf-Nband, graf_thesis}. The indices $i$, $j$, and $l$ belong 
to $\{x, y, z\}$ and label Cartesian components of the three-dimensional 
vectors and tensors throughout. Evaluating Eq.~\eqref{eqomm} for 
$\mathcal{H}(\delta\bs{k})$ yields
\begin{align}
\label{eqbcomm}
\bs{\Omega}_s(\bs{k}) = \frac{(-1)^{s+1}\,v_z\,v_0\,\Delta}
{2\,\epsilon^3\,k_\perp}
\left\{k_y,\,-k_x,\,0\right\}, \quad
\boldsymbol{m}_s(\bs{k}) = \frac{- \,e\,v_z\,v_0\,\Delta}
{2\,\epsilon^2\,k_\perp}
\left\{k_y,\,-k_x,\,0\right\}.
\end{align}
While the BC changes sign between the two bands, the OMM does not, and 
we therefore drop the band label $s$ from $\boldsymbol{m}_s(\bs{k})$ 
henceforth. Decomposing the 3d system into a family of 
2d slices labelled by the toroidal angle $\phi$, each 
subsystem contains two antipodal points related by time-reversal 
symmetry ($\mathcal{T}$), which carry opposite signs of the BC and the OMM:
$\bs{\Omega}_s(\kappa=0,\phi+\pi) = -\, \bs{\Omega}_s(\kappa=0, \phi)$.
This is caused by the planar-vortex structure of $\bs \Omega_s $- and ${\boldsymbol{m}}$-fields.

\begin{figure}[t!]
\centering 
\subfigure[]{\includegraphics[width=0.23 \textwidth]{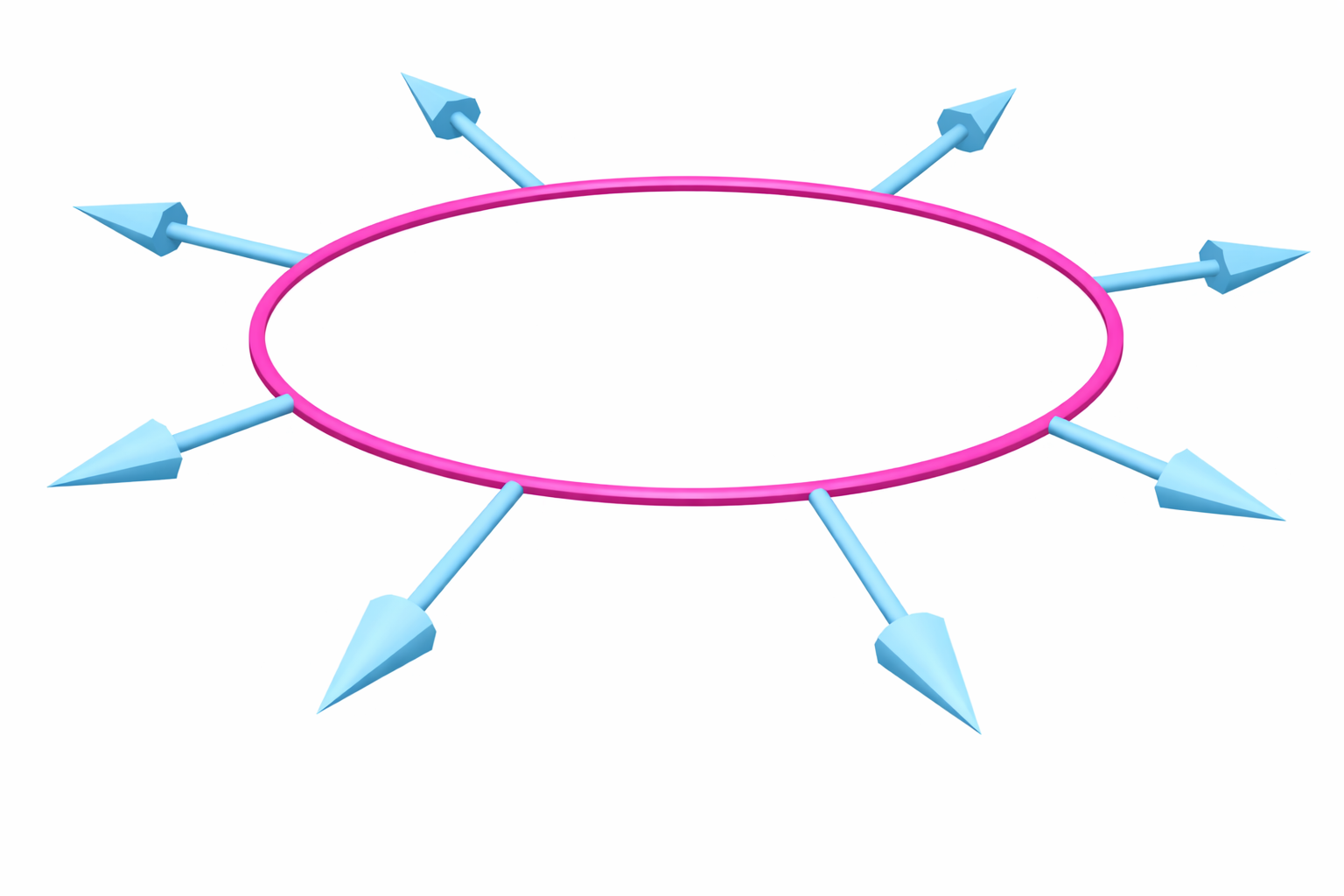}}\hspace{0.5 cm}
\subfigure[]{\includegraphics[width=0.24 \textwidth]{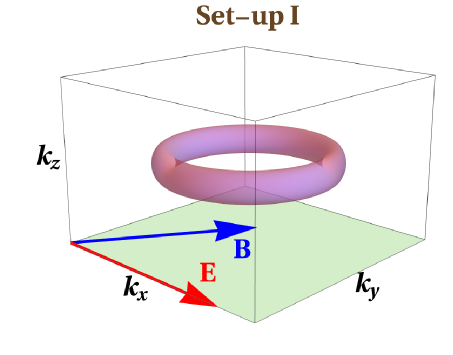}\hspace{-0.25 cm}
\includegraphics[width=0.24\textwidth]{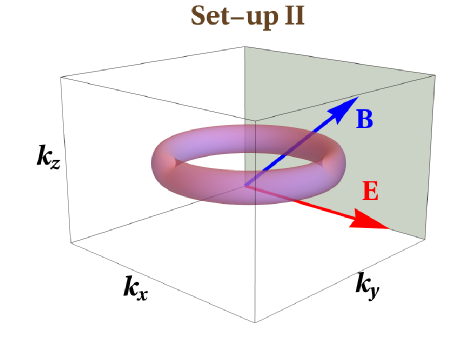}\hspace{-0.25 cm}
\includegraphics[width=0.24 \textwidth]{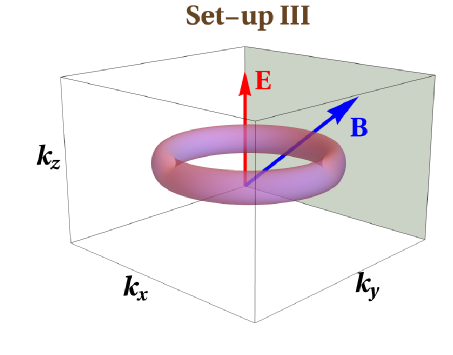} }
\caption{(a) Expansion of the radius of a nodal ring due to the effect of strain.
(b) Schematics of the three configurations used to investigate the planar-Hall
effect in GNRs, illustrating the relative orientations of the 
applied uniform electric (red arrow) and magnetic (blue arrow) fields. 
The three scenarios are labelled as set-up I, set-up II, and set-up III, 
respectively. In each case, the plane containing $\bs{E}$ and $\bs{B}$ 
--- which subtend an angle $\theta$ with respect to one another --- is 
highlighted by a background shading. The coordinate system is chosen 
so that the major radius of the toroidal Fermi surface (shown in light-purple for reference) of the GNR lies in the $k_z = 0$ plane.
\label{figsetup}}
\end{figure}

The BC and the OMM enter the linear-response coefficients through the 
following quantities:
\begin{align}
\label{eqtopo}
& \mathcal{E}_s(\bs{k}) = \varepsilon_s(\bs{k}) 
+ \varepsilon^{\rm (m)}(\bs{k})\,, \quad
\varepsilon^{\rm (m)}(\bs{k}) = -\bs{B}\cdot\boldsymbol{m}(\bs{k})\,, \quad
\boldsymbol{v}_s(\bs{k}) \equiv \nabla_{\bs{k}}\mathcal{E}_s(\bs{k})
= \boldsymbol{v}^{(0,s)}(\bs{k}) + \boldsymbol{v}^{\rm (m)}(\bs{k})\,, 
\nonumber \\
& \boldsymbol{v}^{\rm (m)}(\bs{k}) = \nabla_{\bs{k}}\varepsilon^{\rm (m)}(\bs{k})\,, 
\quad
D_s(\bs{k}) = \left[1 + e\left\{\bs{B}\cdot\bs{\Omega}_s(\bs{k})
\right\}\right]^{-1},
\end{align}
when a magnetic field $\bs B $ is applied.
Here, $\varepsilon^{\rm (m)}$ is the Zeeman-like energy correction induced 
by the OMM \cite{xiao_review, sundaram99_wavepacket, arovas, 
graf_thesis}, $\boldsymbol{v}_s$ is the quasiparticle band-velocity 
renormalized by $\varepsilon^{\rm (m)}$, and $D_s$ is the phase-space 
volume modification factor arising from a nonzero BC \cite{arovas, 
ips-kush-review}. Since the OMM shifts the bare dispersion 
$\varepsilon_s$ to $\mathcal{E}_s$, the shape of the Fermi surface is 
modified accordingly, as illustrated schematically in Fig.~\ref{figfs} 
for the case in which $\bs{B}$ lies within the nodal-loop's plane: the 
originally toroidal Fermi surface deforms into a ring cyclide. If 
$|\bs{B}|$ is increased to a critical threshold, a topological Lifshitz 
transition drives the Fermi surface into a horn cyclide geometry, 
pinching off at a point \cite{yang1, yang_review_nlsm}. We also note 
that the $\varepsilon^{\rm (m)}$ term acts in the same manner as a spectral 
tilt.

Let us work out an estimate of the magnitude of the magnetic fields below which we would avoid a Lifshitz transition. From Eq.~\eqref{eqbcomm}, the OMM magnitude, $|\boldsymbol m(\bs k)|=e\,v_z\,v_0\,\Delta/(2\,\epsilon^2)$, on the (unperturbed) Fermi surface is $\phi$-independent. For an in-plane field $\bs B=B_x\,\hat{\bs x}+B_y\,\hat{\bs y}$ (set-up~I), the Zeeman-like shift $\varepsilon^{\rm(m)}(\bs k)=-\bs B\cdot\boldsymbol m(\bs k)$ therefore has a $\phi$-dependent sign but a fixed magnitude, and its extremum over the ring is $|\varepsilon^{\rm(m)}|_{\rm max}\simeq |\bs B|\,e\,v_z\,v_0\,\Delta/(2\,\mu^2)$, where we have set $\epsilon\to\mu$ to leading order. Since $\mathcal E_s(\bs k)=\varepsilon_s(\bs k)+\varepsilon^{\rm (m)}(\bs k)$, the local Fermi-surface condition at the toroidal angle $\phi$ is $\varepsilon_s(\bs k)=\mu-\varepsilon^{\rm(m)}(\phi)$. The tube of the torus pinches off (i.e., the local Fermi momentum $\kappa_F(\phi)\to0$, so that $\varepsilon_s\to\Delta$) at whichever $\phi$ the shift is most negative. Equating $\Delta=\mu-|\varepsilon^{\rm(m)}|_{\rm max}$ gives the leading-order critical-field estimate for the ring-to-horn-cyclide Lifshitz transition,
\begin{align}
\label{eqBc}
B_c \;\simeq\; \frac{2\,\mu^2\,(\mu-\Delta)}{e\,v_z\,v_0\,\Delta}\,.
\end{align}
This shows explicitly that $B_c$ grows with the chemical potential (as $\sim\mu^3/\Delta$ for $\mu\gg\Delta$), diverges as the gap $\Delta\to 0$ (reflecting that this particular OMM-driven pinch-off mechanism, which is $\propto\Delta$ through $\boldsymbol m$, switches off continuously as the ring becomes gapless). We also observe that $B_c$ is suppressed by larger values of  $v_0$ and $ v_z$ (i.e., by a stiffer, less compressible dispersion). We have verified numerically, using the parameter values of Table~\ref{tab-params}, that $B_c$ obtained from Eq.~\eqref{eqBc} indeed lies well above the field range used to generate our plots, confirming self-consistently that we remain in the ring-cyclide regime throughout.

\subsection{Pseudomagnetic fields}

Subjecting a GNR to a nonuniform lattice strain, we obtain an effective gauge field of the form \cite{nlsm-strain, pikulin-gauge-nlsm, nlr-acoustic}
\begin{align}
\boldsymbol{ A_5} (k_x, k_y, z)= \frac{B_5 \,z} {k_0 } \left(k_x \, {\bs{\hat x}} + k_y \,{\bs{\hat y}}\right )
= B_5 \,z\left( \cos \phi \, {\bs{\hat x}} + \sin \phi \,{\bs{\hat y}}\right ).
\end{align}
This can be achieved by modulating the strength of nearest-neighbour hoppings in the microscopic lattice along the $z$-direction. Depending on the sign of $B_5$, the applied strain concentrically shrinks or expands the NR, as depicted in Fig.~\ref{figsetup}(a). The resulting pseudomagnetic field is given by
\begin{align}
\boldsymbol{ B_5} = \nabla_{\bs r} \cross \boldsymbol{ A_5} = B_5 \left(- \sin \phi  \,{\bs{\hat x}}
+ \cos \phi  \,{\bs{\hat y}} \right) .
\end{align}
Hence, the couplings generated by $ \bs {B_5} $ are chiral. Incorporating the effect of this strain-induced pseudomagnetic field, we define the net effective magnetic field as
\begin{align}
{\bs B}^{\rm{tot}} = \bs B + \bs{ B_5} \,,
\end{align}
which we will use to compute the magnetoelectric conductivity. The primary reason for which we want to investigate the role of $\bs B_5$ is that it couples with \textit{antipodal points} of the NR with \textit{opposite signs}, while $\bs B$ lacks that feature.
Thus the axial field \emph{co-rotates} with the BC/OMM's vector field in a plane perpendicular to the $k_z$-axis, locked to the same vortex structure --- hence
their dot product is a nonzero constant everywhere.

\section{Magnetoelectric conductivity}
\label{secsigma}

In this section, we review the explicit forms of the magnetoconductivity 
tensors arising from a relaxation-time approximation (RTA) of the semiclassical 
Boltzmann equations, in the presence of both the BC and the OMM \cite{ips-kush-review, ips_rahul_ph_strain, rahul-jpcm, ips-ruiz, ips-rsw-ph, ips-shreya}. The 
analysis is valid in the regime $|\bs{B}^{\rm{tot}}| \ll \mu^2/(e\,v_0^2)$, 
which ensures that the inter-Landau-level spacing remains negligible 
compared to other energy scales, justifying a treatment of the dispersion 
as continuous rather than quantised. Equivalently, this condition 
guarantees that the Fermi momentum $\kappa_F$ of the GNR satisfies 
$\kappa_F\,\ell_{B^{\rm{tot}}} \gg 1$, where 
$\ell_B \equiv 1/\sqrt{e\,|\bs{B}^{\rm{tot}}|}$ is the magnetic length 
familiar from the quantum Hall context.

In the weak-field limit, one has 
$e\,|\bs{B}^{\rm{tot}}\cdot\bs{\Omega}_s| \ll 1$, and we work 
consistently to order $\mathcal{O}(|\bs{B}^{\rm{tot}}|^2)$ throughout. 
The phase-space factor $D_s$ is accordingly expanded as
\begin{align}
D_s = 1 - e\left(\bs{B}^{\rm{tot}}\cdot\bs{\Omega}_s\right)
+ e^2\left(\bs{B}^{\rm{tot}}\cdot\bs{\Omega}_s\right)^2
+ \mathcal{O}(|\bs{B}^{\rm{tot}}|^3)\,.
\label{Exp_D}
\end{align}
The same weak-field condition also ensures that the OMM energy 
correction is small relative to the bare dispersion, 
$|\varepsilon^{\rm (m)}(\bs{k})| \ll |\varepsilon_s(\bs{k})|$, since
\begin{align}
\label{eqappr}
|\bs{B}^{\rm{tot}}\cdot\boldsymbol{m}| \equiv 
e\,|\varepsilon_s|\,|\bs{B}^{\rm{tot}}\cdot\bs{\Omega}_s| 
\ll |\varepsilon_s|\,.
\end{align}
This hierarchy permits a Taylor expansion of the derivative of the 
Fermi-Dirac distribution, $f_0(\mathcal{E}_s, \mu, T) \equiv 
\left(1 + e^{(\mathcal{E}_s - \mu)/T}\right)^{-1}$, where $\mu$ is 
the chemical potential and $T$ is the temperature. Retaining terms 
through quadratic order in $|\bs{B}^{\rm{tot}}|$, we obtain
\begin{align}
f_{\rm prime}(\mathcal{E}_s) \equiv 
\frac{\partial f_0(\mathcal{E}_s)}{\partial\mathcal{E}_s}
= f_0^\prime(\varepsilon_s)
+ \varepsilon^{\rm (m)}\,f_0^{\prime\prime}(\varepsilon_s)
+ \frac{1}{2}\left(\varepsilon^{\rm (m)}\right)^2 f_0^{\prime\prime\prime}(\varepsilon_s)
+ \mathcal{O}(|\bs{B}^{\rm{tot}}|^3)\,,
\label{Exp_f}
\end{align}
where the $\mu$- and $T$-dependence has been suppressed for notational 
clarity, and a prime denotes differentiation of $f_0(u)$ with respect 
to its argument $u$.

The electric conductivity $\sigma$ is computed within the semiclassical 
Boltzmann framework, valid at weak magnetic fields, and simplified by 
adopting a momentum-independent relaxation time $\tau$. This 
relaxation-time approximation replaces the full scattering integral by 
a phenomenological rate $\sim 1/\tau$. The scattering lifetime, $\tau $,
drives the fermionic distribution toward a steady state, allowing all 
time-dependence to be dropped from the kinetic 
equation when the external fields and the inhomogeneous strain are static. The detailed derivation of the linearised Boltzmann equation and its solution can be found in Appendix~\ref{appboltz}, which follow our earlier works \cite{ips-kush-review, ips_rahul_ph_strain, rahul-jpcm, 
ips-ruiz, ips-rsw-ph, ips-shreya}.

For a given configuration of the electromagnetic fields, we define the 
planar components of $\sigma$ as those lying in the plane spanned by 
$\bs{E}$ and $\bs{B}$. These decompose naturally into a 
component longitudinal to $\bs{E}$ and an in-plane transverse component, 
referred to as the longitudinal magnetoconductivity (LMC) and the 
planar-Hall conductivity (PHC), respectively. Their explicit forms are 
presented below:
\begin{enumerate}

\item The generic expression for the in-plane components of the magnetoelectric conductivity tensor, contributed by the band with index $s$, is given by 
\begin{align}
\bar \sigma^s _{i j} &= - \, e^2 \,  \tau   
\int \frac{  d^3 \bs k } { (2\, \pi )^3  } \,  
D_s   \,  \left[ \left( v_s  \right)_i + e  \, 
 (  {\boldsymbol  v}_s  \cdot \bs{\Omega }_s  ) \,  B^{\rm{tot}}_{i} \right] 
 \left[ \left( v_s  \right)_j + 
 e  \,  (  {\boldsymbol  v}_s  \cdot \bs{\Omega }_s  ) \,  B^{\rm{tot}}_{j} \right] 
\frac{\partial f_0 (\mathcal{E}_s ) }
{ \partial \mathcal{E}_s   }  .
\label{eq_elec}
\end{align}
For the ease of calculations, we decompose it as $ \bar \sigma^s _{i j} = \sigma_{ i j}^{(s,1) }
+ \sigma_{ i j}^{(s,2) } + \sigma_{ i j}^{(s,3) } + \sigma_{ i j}^{(s,4) }$, where
\begin{align}  
\label{eq_sig_4parts}
 & \sigma_{ i j}^{ (s,1) } = \tau \, e^2 
  \int \frac{d^3   {\bs k}} {(2  \, \pi)^{3}} \,I_{1ij} \,,\quad
\sigma_{ij}^{(s,2)}   =
  \tau \, e^4
\int \frac{d^3   {\bs k}}{(2  \, \pi)^{3}} \,I_2 \,  B^{\rm{tot}}_i \, B^{\rm{tot}}_j \,,
\nn &  \sigma_{ij}^{(s,3)}   =  
{\tau \, e^3}   
 \int \frac{d^3  \bs  k}{(2  \, \pi)^{3}} \,I_{3i} \, B^{\rm{tot}}_j \,, 
 \quad  \sigma_{ij}^{(s,4)}   = 
{\tau \, e^3  }
  \int \frac{d^3   {\bs k}}{(2  \, \pi)^{3}} \,I_{3j}\, B^{\rm{tot}}_i \,,
\nn & I_{1ij} = -\,  D_s   \,
   \left(  v_s   \right)_i \left(  v_s  \right)_j \,
f_0^\prime (\mathcal E_s )\,, \quad
  I_2 = - \, D_s   \left (  {\boldsymbol v}_s     \cdot  
  {\bs  \Omega}_s   \right )^2 
 f_0^\prime (\mathcal E_s )\,, \quad 
 I_{3i} = -\,   D_s   \, 
\left(  v_s  \right)_i  \left ( {\boldsymbol v}_s  
   \cdot   {\bs  \Omega}_s   \right )
  f_0^\prime (\mathcal E_s ) \,.
\end{align}
For the sake of simplicity, we will work in the $T \rightarrow 0 $ limit, such that $f_0^\prime (\mathcal E_s) \rightarrow -\, \delta (\mathcal E_s - \mu  )$.
We note that the results for $T>0$ can be easily obtained by using the relation given by \cite{mermin}
\begin{align}
\label{eqsigmat}
\sigma^s_{ij} (T) =  -\int_{-\infty}^\infty
\sigma^s_{ij} (T=0)\,  \frac{ \partial
 f_{0} ( \mathcal{E}_s,\mu,T   )}
{\partial {\mathcal{E}_s}}  \,.
\end{align}
Up to $\order{ |\bs{B}^{\rm{tot}}|^2}$, we find that \cite{ips-tilted}
\begin{align}
\label{i1}
I_{1ij} & =  \left \lbrace  v^{(0, s)}_{i} \, v^{(0, s)}_{j} 
+  
 v^{(0, s)}_{j} \, v_{i}^{\rm (m)} + v^{(0, s)}_{i} \, v_{j}^{\rm (m)} 
- e \, v^{(0, s)}_{i}  \,v^{(0, s)}_{j}
 \left( \bs{B}^{\rm{tot}} \cdot \bs \Omega_s  \right )
\right \rbrace \delta  (\varepsilon_s - \mu) \nn
 &  \quad +  \varepsilon^{\rm (m)}
\left \lbrace 
v^{(0, s)}_{i}  \, v^{(0, s)}_{j} 
- e \, v^{(0, s)}_{i} \,  v^{(0, s)}_{j}  
\left( \bs{B}^{\rm{tot}} \cdot \bs \Omega_s  \right )
+ v^{(0,s)}_{ j} \, v_{i}^{\rm (m)} + v^{(0, s)}_{i}  
\, v_{j}^{\rm (m)} \right \rbrace
\delta^\prime (\varepsilon_s - \mu)
 \nn & \quad 
 + \left \lbrace 
  e  \, v^{(0, s)}_{i}  \left( \bs{B}^{\rm{tot}} \cdot \bs \Omega_s  \right )
  - v_{i}^{\rm (m)} \right \rbrace 
\left \lbrace e \, v^{(0, s)}_{j} 
 \left( \bs{B}^{\rm{tot}}  \cdot \bs \Omega_s  \right ) -  v_{j}^{\rm (m)} 
 \right \rbrace \,  \delta  (\varepsilon_s - \mu)
+ \frac { v^{(0, s)}_{i} \,v^{(0, s)}_{j} \left( \varepsilon^{\rm (m)} \right)^2 
 \delta^{\prime \prime } (\varepsilon_s - \mu)} 
{2} \,,\nn
I_2 & = \left( \boldsymbol v^{(0, s)} \cdot \bs \Omega_s  \right)^2  
\delta  (\varepsilon_s - \mu)\,,\nn
I_{3i} &= \left [  
\left( \boldsymbol v^{(0, s)} \cdot \bs \Omega_s  \right )
\left  \lbrace
 v_{i}^{\rm (m)} +  v^{(0, s)}_{i} 
- e \,   v^{(0,s)}_{i} \left( \bs{B}^{\rm{tot}} \cdot \bs \Omega_s  
 \right) \right \rbrace
+ v^{(0, s)}_{i}  \left( \boldsymbol v^{\rm (m)} 
\cdot \bs \Omega_s  \right ) \right] 
\delta  (\varepsilon_s - \mu )
 \nn & \qquad 
+ v^{(0, s)}_{i} \,\varepsilon^{\rm (m)} 
\left( \boldsymbol v^{(0, s)} \cdot \bs \Omega_s  \right )
 \delta^\prime (\varepsilon_s - \mu)\,.
\end{align}

\item
The out-of-plane components are captured by the so-called anomalous-Hall part (denoted by $\sigma^{\text{(ah)}}_s $) and the Lorentz-force-operator contributions \cite{ips-rsw-ph, ips_tilted_dirac, ips-spin1-ph}. Expanding up to $\order{ |\bs{B}^{\rm{tot}}|^3}$, we find that
\begin{align}
\label{AH}
(\sigma^{\text{(ah)}}_s)_{ij} & = - \, e^2 \,\epsilon_{ijl}
\int \frac{  d^3 {\bs k} }  {(2 \,\pi )^{3} } \, \Omega_s^l  
\left [ f_0 ( \varepsilon_s  ) 
+  \varepsilon^{ (m) } \,  
f'_0 (\varepsilon_{ s} ) + \frac{1}{2} \left(  \varepsilon^{\rm (m)} \right)^2  \,
  f^{\prime \prime}_0 ( \varepsilon_s  )   +  \frac{1}{6} \, 
\left(  \varepsilon^{\rm (m)} \right)^3  \,
f_0^{ \prime \prime \prime} (\varepsilon_s ) + \order{ |\bs{B}^{\rm{tot} }|^4}  \right ].
\end{align}
Clearly, the first term is independent of $\bs{B}^{\rm{tot} }$ (which is the origin of the nomenclature of ``anomalous Hall''), and it vanishes identically in our system. The nonzero terms appear only when we correctly account for the OMM-part, thus showing the importance of not omitting the OMM-contributions. Next, we have
\begin{align}
\label{lf_cond1}
\left( \sigma_s^{\rm{(lf)}} \right)_{ij} &= -e^2 \tau \int \frac{d^3\bs{k}}{(2\pi)^3} \left[ \left(v_s\right)_i 
+ \left(u_s\right)_i \right] f_0^\prime( \mathcal{E}_s) \frac{\partial \mathcal{Y}_s}{\partial E_j}\,,\nn
{\boldsymbol u}_s & = e  \, 
 (  {\boldsymbol  v}_s \cdot \bs{\Omega }_s  ) \,  \bs{B}^{\rm{tot} }, \quad
 \mathcal{Y}_s = \sum_{n=1}^{\infty} \left(e \,\tau\, 
D_s\right)^n \check{L}^n \left[  D_s \left\{ {\boldsymbol  v}_s +{\boldsymbol  u}_s \right\} \cdot \bs{E} \right],
\text{ and } \check{L} =\left ({\boldsymbol  v}_s \times \bs{B}^{\rm{tot} } \right ) \cdot \nabla_{\bs{k}} \,.
\end{align}
The symbol $\check{L}$ denotes the so-called Lorentz-force (LF) operator and, comprising a differentiation operator (with respect to $ \bs k$), it acts on everything appearing on its right-hand side. The detailed derivation of the expressions can be found in Refs.~\cite{ips_tilted_dirac, ips-spin1-ph}. We note that the action of $\check{L}$ each time contributes to one linear power of
$ |\bs{B}^{\rm tot} |$, while additional powers enter through $ \boldsymbol{v}^{\rm (m)}$ and the series-expansions of $D_s$ and $f'_0(\mathcal{E}_s)$ in $  |\bs{B}^{\rm tot} |$. Following the convention of Ref.~\cite{ips-spin1-ph}, we label each contribution to the integral for computing the current-density as $\bs {\bs{\mathcal{ N }}}_{n,p}$, where the subscript $n$ indicates the $n^{\rm th}$ term in the summation, ${\mathcal Y}_s $ (where $\check{L}^n $ generates $ |\bs{B}^{\rm tot}|^n$). The subscript $p$ in $\bs {\bs{\mathcal{ N }}}_{n,p}$ indicates that overall $ |\bs{B}^{\rm tot} |^p$ appears in that part. More details can be found in Appendix~\ref{appLF}.

\end{enumerate}

\begin{table}[t!]
\centering
\begin{tabular}{|c|c|c|}
\hline
 Parameter &     Natural Units  \\ \hline
$v_0$ from Ref.~\cite{phe_nlsm} & $0.0004$\\  
\hline
$v_z$ from Ref.~\cite{phe_nlsm} &  $0.00045$
\\  
\hline
$k_0$ from Refs.~\cite{phe_nlsm, pikulin-gauge-nlsm} &  $ 160 $  \text{eV} 
\\
\hline
$\Delta$ from Ref.~\cite{phe_nlsm} &  $0.02$  \text{eV}
\\ \hline
$\tau$ from Ref.~\cite{yang1} &  $15.2  $ eV$^{-1}$  
\\  \hline
$ B $ from Ref.~\cite{phe_nlsm} &  $ 0 $ --- $ 100 $ eV$^2 $ \\ \hline
$ B_5 $ from Ref.~\cite{pikulin-gauge-nlsm} & $ 0$ --- $ 6.25 $ eV$^2$ \\ \hline 
$\mu$ from Ref.~\cite{phe_nlsm} &  $0.04 $ eV  \\ \hline
\end{tabular}
\caption{\label{tab-params}The ranges of values for the various parameters, chosen to generate representative plots of the conductivity tensors, are tabulated here. We use natural units and, hence, set $\hbar= c = k_B = e = 1$. However, we retain the symbol $e$ in the analytical expressions for the sake of book-keeping.}
\end{table}

While deriving conductivity, we will assume that a positive chemical potential $\mu$ is applied (i.e., $\mu>\Delta$). Hence, we will do all the calculations for the conduction band (i.e., we set $s = 2$), and we will employ the coordinate transformations shown in Eq.~\eqref{eqtrs} to perform the integrations. We will drop the band-index and divide up the final expression for in-plane components as
\begin{align}
\bar \sigma_{ij} = \sigma_{ij}^{\rm (d)} + \sigma_{ij}^{\rm (bc)} +\sigma_{ij}^{\rm (m)} + \sigma_{ij}^{({\rm conc})}\,,
\end{align}
where the superscripts of ``(d)'', ``(bc)'', ``(m)'', and ``(conc)'' are used to denote the Drude, BC-only, OMM-only, and joint BC-OMM-contributed parts (i.e., concurrent), respectively. The Drude part is the one which is independent of the applied magnetic field, and is nonzero only for the longitudinal components [i.e., $ \sigma_{ij}^{\rm (d)} \propto \delta_{ij} $]. The BC-only part does not contain any contribution from the OMM and, therefore, survives even when OMM is not included. In a similar spirit, we  divide up the final expression for Lorentz-force-induced components as 
\begin{align}
\sigma^{\text{(lf)}}_{ij} &= \sigma^{\text{lf,(h)}}_{ij} + \sigma^{\text{lf,(bc)}}_{ij} +\sigma^{\text{lf,(m)}}_{ij} 
+ \sigma^{\text{lf,(conc)}}_{ij} \,.
\end{align}
$\sigma^{\text{lf,(h)}}$ contains the classical Hall-conductivity part.

In the following, we introduce the variable,
\begin{align}
\zeta = \sqrt{k_0^2\, v_0^2 + \Delta^2 - \mu^2}\,,
\end{align} 
to shorten various expressions.
The ranges of the values of the parameters in some realistic scenarios have been shown in Table
\ref{tab-params}, which we use for plotting the features contained in our final analytic expressions.
The magnitude of $B_5$ can be estimated from the microscopic treatment of strain carried out in Ref.~\cite{pikulin-gauge-nlsm}. There, $B_5^{-1}  = Q\, R\, a $, where $Q = k_0 \, a$, $R = 8000\, a$, $a \approx 1$ nm (in SI units) $\equiv 1/200 $ eV$^{-1}$ (in natural units) --- this yields $B_5 \approx 6.25  $ eV$^2$. For plotting the conductivity curves, we define
\begin{align}
\Delta \sigma_{ij}^{(\delta)} =\frac{ \sigma_{ij}^{(\delta)} }
{\sigma_{ij}^{(\delta)} \vert_{B_5 = 0 }} -1\,.
\end{align}

\subsection{Set-up I: 
\texorpdfstring{$\bs{E}=E_x\, {\bs{\hat x}}$}{E along x} and 
\texorpdfstring{$ \bs{B}=B_x \,{\bs{\hat x}} + B_y \,{\bs{\hat y}} $}{B field}}
\label{secset1}

\begin{figure}[t!]
\centering 
\subfigure[]{\includegraphics[width= 0.75 \textwidth]{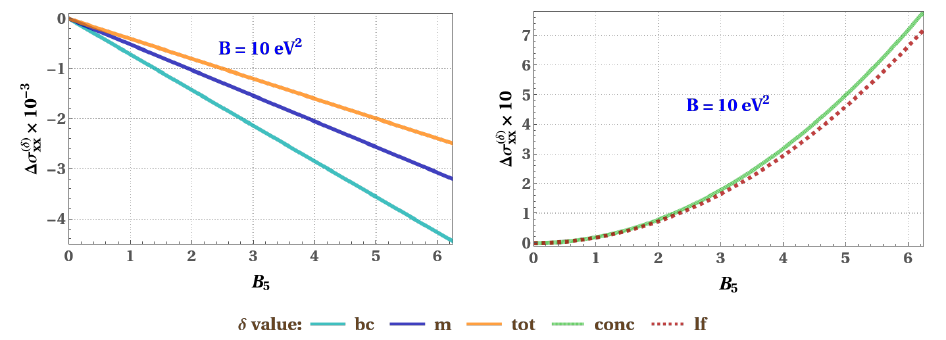}}\\
\subfigure[]{\includegraphics[width= 0.35 \textwidth]{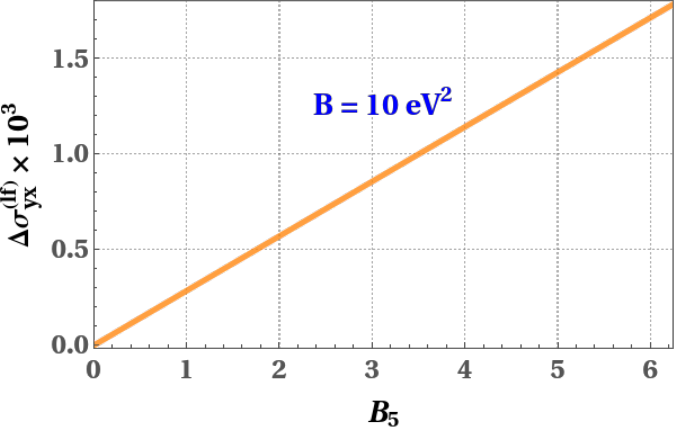}} \hspace{ 0.5 cm}
\subfigure[]{\includegraphics[width= 0.35 \textwidth]{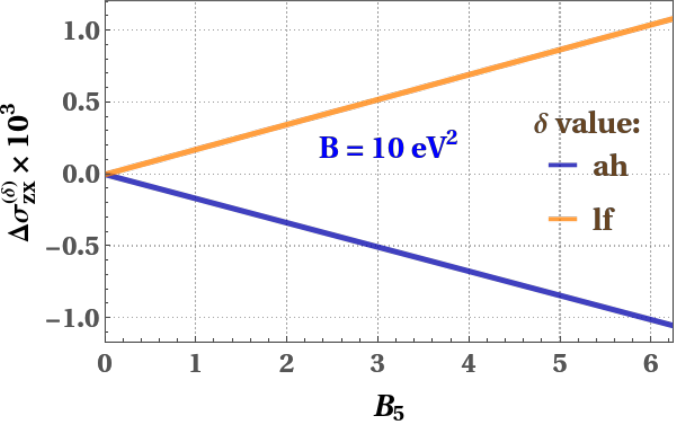}}
\caption{Set-up I: Behaviour of (a) $ \Delta \sigma_{xx}^{(\delta)} $, (b) $\Delta \sigma^{\rm(lf)}_{yx}$, and (c) $\Delta \sigma^{(\delta)}_{zx}$ as functions of $B_5$ (in eV$^2$), taking into account various contributions. The individual longitudinal
contributions from the BC, OMM, their concurrent effects, and $\bar \sigma_{xx}$ (labelled as $\sigma_{xx}^{\rm(tot)}$) are shown separately. The anomalous-Hall and LF-induced parts are also included. We have used the parameter values from Table~\ref{tab-params} and set $\theta = \pi/4$.
\label{figset1}}
\end{figure}

In the set-up I shown in Fig.~\ref{figsetup}(b), we have $\bs{E} = E_x\, {\bs{\hat x}}$ and $ \bs{B}=B_x \,{\bs{\hat x}} + B_y \,{\bs{\hat y}} $. Consequently, Eq.~\eqref{eqtopo} translates into  $ \varepsilon^{\rm (m)} (\bs{k}) 
 = \frac{e\,v_z\, v_0\, \Delta} {2\, \epsilon^2}\, 
\left[ \frac{\left( B_x-B_5 \sin{\phi} \right)\, k_y - \left( B_y+B_5 \cos {\phi} \right)\, k_x } {k_{\perp}} \right]$. The full expression for the OMM-induced velocity-correction is shown in Appendix~\ref{appvel}. Using these, Eq.~\eqref{i1} leads to
\begin{align}
\label{eqsigxxset1}
\sigma^{\rm (d)}_{xx} & = \frac{\tau \, e^2 \, k_0 \, v_0 }
{8 \, \pi \, v_z \, \mu}  
\left(\mu^2 - \Delta^2 \right), \quad
\sigma^{\rm (bc)}_{xx}  =
\frac{\tau \, e^4 \, v_z \, v_0^3  \, \Delta^2 \, k_0 \left( \mu^2 - \Delta^2 \right)}
{128 \, \pi \, \mu^7} 
\left[ 
- \, \frac{8\, B_5\,\mu^3}{e\,v_0\,v_z\,\Delta}
+\left( B_x^2 + 3 \, B_y^2 + 4\,B_5^2\right)  \right],
\nn \sigma^{\rm (m)}_{xx} &   = 
\frac{   \tau \, e^4 \,v_z \, v_0^3 \, \Delta^2}
{128 \, \pi \, \mu^7} 
 \left[  \frac{\,8\,k_0\,B_5\,\mu^3\, ( 3\,\mu^2 -\Delta^2)}{\,e\,v_0\,v_z\,\Delta}
- \left(B_x^2 + 3 \, B_y^2 + 4\, B_5^2 \right)  k_0  \left( 6\, \mu^2 - 5\, \Delta^2 \right) 
+\frac{2 \left(B_x^2 + 3 \, B_y^2 \right) \mu^4}  {v_0\, \zeta}
\right] , \quad
\nn \sigma^{({\rm conc})}_{xx} &=\,\frac{-\,\tau \, e^4 \, v_z \, v_0^3  \, \Delta^4 \, k_0 \left( B_x^2 + 3 \, B_y^2 + 4\,B_5^2\right)
  } {64 \, \pi \, \mu^7}\,,
  \end{align}
\begin{align}
\label{eqsigyxset1}
\sigma^{\rm (d)}_{yx} & = 0 \,,
\quad
\sigma^{\rm (bc)}_{yx}  =
\frac{ -\, \tau \, e^4 \, v_z \, v_0^3  \, \Delta^2 \, B_x \, B_y}
{ 64 \, \pi \, \mu^7} 
\, k_0 \left(  \mu^2 - \Delta^2 \right) ,\nn
\nn \sigma^{\rm (m)}_{yx} &   = 
\frac{   \tau \, e^4 \,v_z \, v_0^3 \, \Delta^2\,  B_x \, B_y }
{64 \, \pi \, \mu^7} 
 \left [  k_0 \left( 6\, \mu^2 - 5\, \Delta^2\right)- \frac{2 \,\mu^4} {v_0\, \zeta} \right ] ,\quad
 \sigma^{({\rm conc})}_{yx}   = 
\frac{   \tau \, e^4 \,v_z \, v_0^3 \, \Delta^4\, k_0\, B_x \, B_y }
{32 \, \pi \, \mu^7}\,,\nn
 \sigma^{\text{(ah)}}_{zx} &= 
\frac{ - \,e^3\, v_z \, v_0 \, k_0 \,\Delta^2 \, B_y} {16 \, \pi\, \mu^4} 
 \left[
 1-\, \frac{3\,B_5\,e\,v_0\,v_z\,\Delta}{\mu^3}
 +\frac{ 9\, e^2  \,v_z^2 \, v_0^2  \, \Delta^2 
\left(B_x^2 + B_y^2 + 4\,B_5^2\right) }
 { 4\, \mu^6 }  \right].
\end{align}
We note that the $\theta $-dependence of $\bar \sigma_{xx}$ comes exclusively from the terms containing $B_x^2$ and $B_y^2$; the $B_5$-
and $B_5^2$-dependent terms merely shift the values in a $\bs B$-independent way.
Interestingly, $\bar \sigma_{yx} $ gets no contribution from strain.
For the $\check L$-induced contribution obtained from Eq.~\eqref{lf_cond1}, 
we expand the summation up to $n=3$. All the terms from the intermediate steps can be found in Appendix~\ref{appset1}.
The overall contributions to the in-plane components turn out to be
\begin{align}
\label{eqset1-lfinplane}
\sigma^{\text{(lf)}}_{xx} & =  \frac{ \tau^3 \, e^4\, v_0^3 \,v_z  }
{ 32 \,\pi \,\zeta \,\mu^4 }
\Big[
 B_5\, e \,v_z \,\Delta  \, \Big\{B_x^2 \left(k_0 \, v_0\, (13\, k_0\,  v_0-16 \, \zeta)
+6 \,(\Delta^2-\mu^2)\right) 
+ B_y^2
\left(k_0 \, v_0(23\, k_0 \, v_0- 32\, \zeta)+ 18\, (\Delta^2-\mu^2)
\right) \Big\}
\nn & \hspace{2.25 cm}
+4\, B_5^2 \,k_0 \,\zeta \,\mu \,(\Delta^2-\mu^2)
- 2 \left (B_x^2+3\, B_y^2 \right ) k_0^2 \,v_0 \,\zeta\, \mu 
\left( k_0 \,v_0 -\zeta\right) -4 \,B_5^3 \,e \,k_0\, v_0 \,v_z \,\Delta\, \zeta
\Big]\,, 
\nn \sigma^{\text{(lf)}}_{yx} & = \frac{ \tau^3  \, e^4 \,  v_z \,v_0^3}
{16 \,\pi\, \zeta \,\mu^4} B_x\, B_y \left[
B_5 \, e\, v_z \,\Delta\left\{k_0 \, v_0 \left (8\, \zeta-5 \, k_0\,  v_0 \right )
-6\,(\Delta^2-\mu^2)\right\}
+ 2\,  k_0^2 \, v_0 \, \zeta \, \mu\left(k_0 \, v_0- \zeta\right)\right].
\end{align}
Adding up all the out-of-plane contributions from $n=1$ and $ n=3$, we get 
\begin{align}
\label{eqsigzsxset1}
\sigma^{\text{(lf)}}_{zx}
&= \frac{\tau^2 \, e^3 \, v_z \, v_0\, B_y} {128\, \pi \, \mu^8\ }\,
\Big[
16\, k_0 \, \mu^6  \left(\Delta^2 - \mu^2\right)
+ 32\, e \, v_z \, \Delta \,\mu^5\,B_5 \left( \zeta -2\, k_0 \, v_0 \right)
- 12 \, \Delta^4 \,e^2 \, k_0 \,v_0^2\, v_z^2 \left(B_x^2+B_y^2+4 B_5^2\right)
\nn &   \hspace{ 2.75 cm }
+ 24  \, e^2 \,  k_0^2  \, \mu^4  \, \tau^2  \, v_0^3  \, v_z^2 \left(B_x^2+B_y^2+16  \, B_5^2\right) 
\left(k_0  \, v_0 - \zeta \right)
+ \frac{ 3  \, \Delta^2 \,  e^2 \,  v_0  \, v_z^2 \,\mu^2 \left(B_x^2+B_y^2\right) 
\left(7  \, \zeta \, k_0  \,  v_0 - 2  \, \mu^2 \right) } {\zeta}
 \nn & \hspace{ 2.75 cm }
-\frac{4 \, B_5^2 \, \Delta^2 \, e^2 \, \mu^2\,  v_0\,  v_z^2 
\left \lbrace 4 \left(\Delta^2+\zeta^2\right) + k_0 \, v_0 \left(4 \, k_0 \, v_0-29 \, \zeta \right)
\right \rbrace} {\zeta}
- 144  \, B_5^2 \,  e^2 \,k_0  \, \mu^4  \, \tau^2  \, v_0^2\, v_z^2 \left(\mu^2-\Delta^2\right)
 \Big]\,.
\end{align}
Some representative curves, for all the contributions discussed above, are illustrated in Fig.~\ref{figset1} as functions of $B_5$. 
For the longitudinal component, $\Delta \sigma_{xx}^{\rm(conc)}$ is of the same order of magnitude as $\Delta \sigma_{xx}^{\rm (lf)}$, which are smaller than $\Delta \sigma_{xx}^{\rm(bc)}$ and $\Delta \sigma_{xx}^{\rm(m)}$ (and, hence, $\Delta \bar \sigma_{xx}$) by orders of magnitude (precisely, smaller by $ \sim 10^{-3} $). 
The transverse components are smaller than the net longitudinal component by order of $10^{-6}$.

\subsection{Set-up II: \texorpdfstring{$\bs{E}=E_x\,  {\bs{\hat x}}$}{E-field}
and \texorpdfstring{$ \bs{B}= B_x \, {\bs{\hat x}} + B_z\, \bs{\hat z} $}{B-field}}
\label{secset2}

\begin{figure}[t!]
\centering 
\subfigure[]{\includegraphics[width= 0.75 \textwidth]{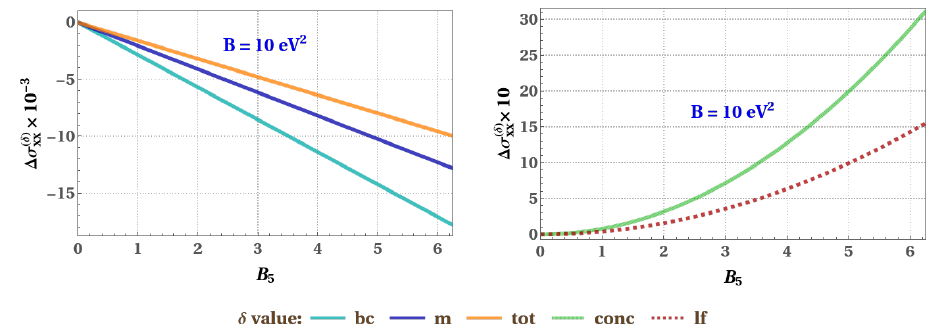}}\\
\subfigure[]{\includegraphics[width= 0.35\textwidth]{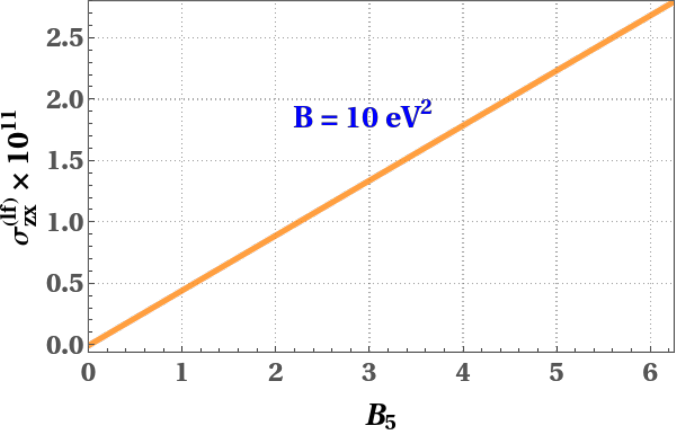}} \hspace{ 0.5 cm}
\subfigure[]{\includegraphics[width= 0.35\textwidth]{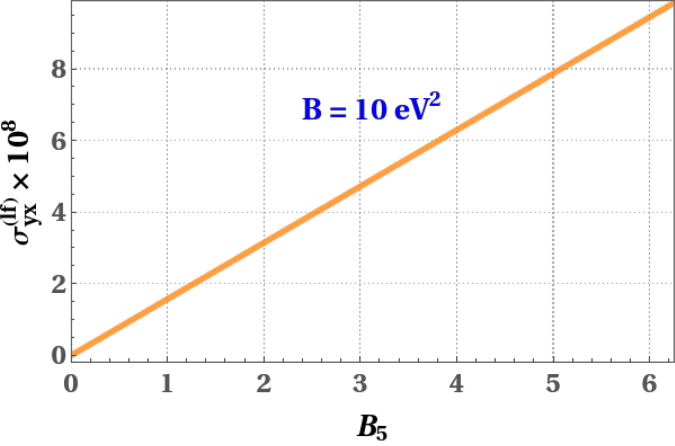}}
\caption{Set-up II: Behaviour of (a) $ \Delta \sigma_{xx}^{(\delta)} $, (b) $ \sigma^{\rm(lf)}_{zx}$ (in eV), and (c) $  \sigma^{\rm(lf)}_{yx}$ (in eV) as functions of $B_5$ (in eV$^2$), taking into account various contributions. The individual longitudinal
contributions from the BC, OMM, their concurrent effects, and $\bar \sigma_{xx}$ (labelled as $\sigma_{xx}^{\rm(tot)}$) are shown separately. The LF-induced parts are also included. There is no nonzero anomalous-Hall part. We have used the parameter values from Table~\ref{tab-params} and set $\theta = \pi/4$.
\label{figset2}}
\end{figure}

In the set-up II shown in Fig.~\ref{figsetup}(b), we have $\bs{E}  = E_x \,{\bs{\hat x}}$ and $\bs{B}= B_x \, {\bs{\hat x}}+B_z\,  {\bs{\hat z}}$. Consequently, Eq.~\eqref{eqtopo} evaluates to 
 $ \varepsilon^{\rm (m)} (\bs{k}) 
  = \frac{e\,v_z\, v_0\, \Delta}
{2\, \epsilon^2}
\left[ \frac{\left( B_x-B_5 \sin{\phi} \right)\, k_y - B_5 \cos {\phi} \, k_x } {k_{\perp}} 
 \right].$
The full expression for the OMM-induced velocity-correction is shown in Appendix~\ref{appvel}. Using these, Eq.~\eqref{i1} leads to
\begin{align}
\label{eqset2long}
\sigma^{\rm (d)}_{xx} & = \frac{\tau \, e^2 \, k_0\, v_0 }
{8 \, \pi \,v_z \,\mu} 
 \left(\mu^2 - \Delta^2 \right),\quad
\sigma^{\rm (bc)}_{xx}  =
\frac{\tau \, e^4  \,v_z\, v_0^3 \, \Delta^2 \, k_0
\left( \mu^2 - \Delta^2 \right) } {128 \, \pi \, \mu^7}\, 
\Big (
-\frac{8\,B_5\,\mu^3}{e\,v_0\,v_z\,\Delta}
+ B_x^2+4\,B_5^2 \Big) \, , \nn
\sigma^{\rm (m)}_{xx} &   = \frac{   \tau \, e^4 \,v_z \, v_0^3 \, \Delta^2}
{128 \, \pi \, \mu^7} 
 \left[ 
\frac{8\,B_5 \,k_0 \,\mu^3 \, ( 3\,\mu^2 - \Delta^2)}
{\,e\,v_0\,v_z\,\Delta}
-(B_x^2 + 4\, B_5^2)\, k_0 \left( 6\, \mu^2 - 5\, \Delta^2 \right)
+\frac{2 \,B_x^2  \,\mu^4} {v_0\, \zeta} 
\right] ,
\nn \sigma^{({\rm conc})}_{xx} & = 
\frac{ -\,\tau\, e^4\, v_z \, v_0^3 \, \Delta^4 \, k_0 \,(B_x^2+4\,B_5^2)}
{64\, \pi\, \mu^7}\,, 
\end{align}
\begin{align}
\bar \sigma_{zx} =  \sigma^{\text{(ah)}}_{yx}  =0\,.
\end{align}
Again, we note that the $\theta $-dependence of $\bar \sigma_{xx}$ comes exclusively from the terms containing $B_x^2$; the $B_5$- and $B_5^2$-dependent terms merely shift the values in a $ B_x^2$-independent way. The in-plane transverse component, $\bar \sigma_{zx}$, is identically zero and no out-of-plane component from the anomalous-Hall integrand survives.
For the $\check L$-induced parts, the individual contributions from the first three terms of the sum are shown in Appendix~\ref{appset2}.
The net in-plane results are as follows:
\begin{align} 
\sigma^{\text{(lf)}}_{xx} &  =\frac{\tau^3 \, e^4 \,v_0^3}{32\,\pi\, \mu^4}
\Bigg[
-4\, B_5^3 \,e \,k_0\, v_0 \,v_z^2 \,\Delta
+4\, B_5^2 \,k_0 \,v_z \,\mu \,(\Delta^2-\mu^2)
\nn &\hspace{ 2 cm}
+\frac{2 \,k_0 \,v_0 \,\mu}{v_z \,\zeta}
\left\{ 2 \,B_z^2\, v_0\,(2 \,k_0^2 \,v_0^2 \left( \zeta- k_0 \, v_0\right)
-\zeta \left( \Delta^2- \mu^2) \right)
+ B_x^2 \,k_0\, v_z^2 \left(k_0 \,v_0 (k_0\, v_0-\zeta)+\Delta^2-\mu^2\right)
\right\}
\nn & \hspace{ 2 cm}
+B_5 \,e \,\Delta
\left\{
\frac{B_x^2\, v_z^2}{\zeta}
\left(k_0 \,v_0 \,(13\, k_0 \,v_0-16\, \zeta)+6 \, (\Delta^2-\mu^2)\right)
+4 \,B_z^2 \,k_0\, v_0^3 \left(
6 -\frac{ k_0\, v_0 \left(7\, \zeta^2 -k_0^2\,v_0^2 \right)}
{\zeta^3}\right)\right\} \Bigg],
\nn \sigma^{\text{(lf)}}_{zx} & = \frac{ \tau^3 \, e^5\,  v_0^3 \,v_z^2\, \Delta\, B_5\, B_x\, B_z }
{ 8\, \pi\, \zeta \,\mu^4 }
\left[5 \,k_0\, v_0 \,(\zeta-k_0 \,v_0)-3 \,(\Delta^2-\mu^2)\right].
\end{align}
The overall out-of-plane contribution is obtained as
\begin{align}
\sigma^{\text{(lf)}}_{yx} & =\frac{\tau^2\, e^4 \, v_0^4 \, B_z 
} {32\, \pi\, v_z\, \mu^6 \, \zeta^3 }
\Big[
4\, B_5 \,v_z \, k_0 \,\Delta
\left( k_0\, v_0 - \zeta \right)
\zeta^2 \,\mu^3
+ 2 \,\Delta^2  \, e  \, k_0 \,  v_0 \left(B_x^2-2 \, B_5^2\right) v_z^2 \left(\Delta^2+k_0^2 \, v_0^2\right) 
\left(k_0  \, v_0-\zeta  \right)
\nn & \hspace{2.75 cm} 
+ \Delta^2  \,e  \,\mu^2 \, B_x^2  \,v_z^2 \left \lbrace 3\, (\Delta^2 -\mu^2 )
+ 2  \,\zeta \,  k_0 \, v_0\right \rbrace
+ 2  \, B_5^2  \, \Delta^2  \, e  \, k_0  \, \mu^2  \, v_0  \, v_z^2 \left(3  \, k_0  \, v_0-2 \,  \zeta \right)
\nn & \hspace{2.75 cm} 
+ 32 \,  e  \, \zeta^2  \, k_0^3  \, \mu^2  \, \tau^2 v_0^3 \left(3 \,  B_5^2-B_x^2\right) v_z^2 
\left(\zeta -k_0 \,  v_0\right) 
-4  \, e \, \zeta^2  \, k_0  \, \mu^2  \, \tau^2  \, v_0 \,  B_x^2 \left(\Delta^2-\mu^2\right) 
v_z^2 \left(\zeta -5 \,  k_0  \, v_0\right)
\nn & \hspace{2.75 cm} 
+ 24  \,B_5^2 \,e  \,\zeta^2  \,k_0  \,\mu^2  \,\tau^2  \,v_0 \,  v_z^2
\left(\Delta^2-\mu^2\right)  \left(\zeta -3 \, k_0  \,v_0\right)
\nn & \hspace{2.75 cm}
-8  \,e \, k_0 \, \mu^2  \,\tau^2  \,v_0^3 \, B_z^2 
\left\lbrace \zeta  \left(\Delta^2-\mu^2\right)^2+k_0^2 \, v_0^2 \left(\Delta^2-\mu^2\right) 
\left(5  \, k_0  \, v_0-3  \, \zeta \right)
+ 4 \, k_0^4 \, v_0^4 \left(k_0  \,v_0-\zeta \right)
   \right \rbrace \Big]\,.
\end{align}
Fig.~\ref{figset2} illustrates the conductivity curves for some representative values of the parameters. For the longitudinal component, $\Delta \sigma_{xx}^{\rm(conc)}$ is of the same order of magnitude as $\Delta \sigma_{xx}^{\rm (lf)}$, which are smaller than $\Delta \sigma_{xx}^{\rm(bc)}$ and $\Delta \sigma_{xx}^{\rm(m)}$ (and, hence, $\Delta \bar \sigma_{xx}$) by orders of magnitude (precisely, smaller by $\sim 10^{-3} $). The transverse components of conductivity get contributions \textit{only} from the LF-induced integrals.

\subsection{Set-up III: 
\texorpdfstring{$\bs{E}=E_z\,\bs{\hat{z}}$}{E along z} and 
\texorpdfstring{$\bs{B}= B_x\, \bs{\hat{x}}+B_z\,\bs{\hat{z}}$}{B with z component}}
\label{secset3}

\begin{figure}[t!]
\centering 
\subfigure[]{\includegraphics[width= 0.75 \textwidth]{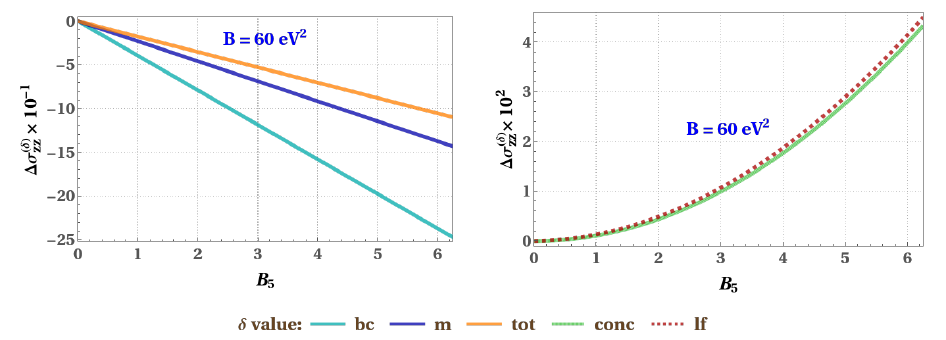}}\\
\subfigure[]{\includegraphics[width= 0.35\textwidth]{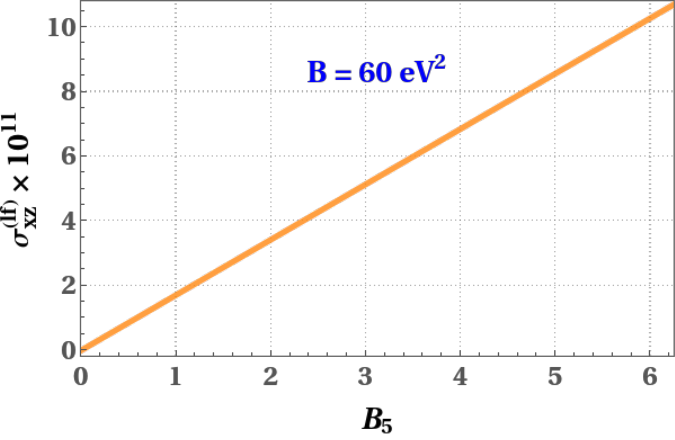}} \hspace{ 0.5 cm}
\subfigure[]{\includegraphics[width= 0.35\textwidth]{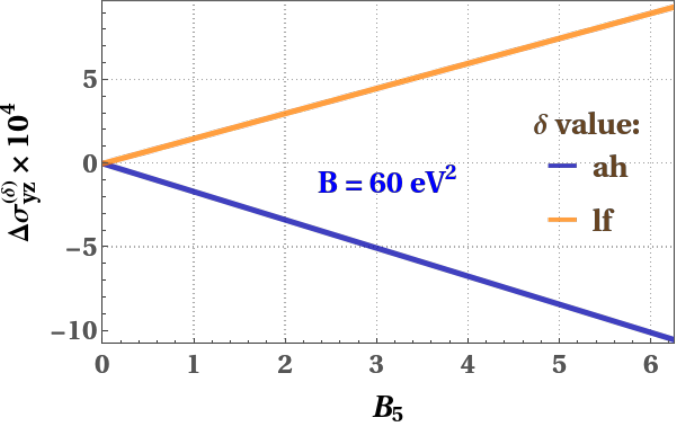}}
\caption{Set-up III: Behaviour of (a) $ \Delta \sigma_{zz}^{(\delta)} $, (b) $ \sigma^{\rm(lf)}_{xz}$ (in eV), and (c) $ \Delta \sigma^{(\delta)}_{yz}$ as functions of $B_5$ (in eV$^2$), taking into account various contributions. The individual longitudinal
contributions from the BC, OMM, their concurrent effects, and $\bar \sigma_{zz}$ (labelled as $\sigma_{zz}^{\rm(tot)}$) are shown separately. The anomalous-Hall and LF-induced parts are also included. We have used the parameter values from Table~\ref{tab-params} and set $\theta = \pi/4$.
\label{figset3}}
\end{figure}

In the set-up III shown in Fig.~\ref{figsetup}(b), we have $\bs{E}= E_z\,{\bs{\hat z}}$ and $\bs{B}= B_x\, \bs{\hat{x}}+B_z\,\bs{\hat{z}}$.
Since the magnetic field is in the same plane as in set-up II, $\varepsilon^{\rm (m)} (\bs{k}) $ and $ \boldsymbol
 v^{\rm (m)} (\bs{k}) $ are the same as in the previous subsection. Using those expressions in Eq.~\eqref{i1}, we obtain
\begin{align}
\label{eqset3long}
\sigma^{\rm (d)}_{zz} & = \frac{\tau \, e^2 \, v_z \, k_0}
{ 4 \, \pi \, v_0 \,\mu } 
 \left(\mu^2 - \Delta^2 \right),\quad
\sigma^{\rm (bc)}_{zz}  = 
\frac{\tau \, e^4  \,v_z^3\, v_0 \, \Delta^2 \, k_0
\left( \mu^2 - \Delta^2 \right) }
{32 \, \pi \, \mu^7}\, \left[ B_x^2+2\,B_5^2-\frac{4\,B_5\,\mu^3}{e\,v_0\,v_z\,\Delta} \right] ,\quad \nn
 \sigma^{\rm (m)}_{zz} & =\frac{\tau \, e^4  \,v_z^3\, v_0 \, \Delta^2 \, k_0
}
{32 \, \pi \, \mu^7}
\left[ \left(B_x^2+2\,B_5^2\,\right)\left(5\, \Delta^2 -6\, \mu^2 \right) +\frac{4\,B_5\,\mu^3\left(3\, \mu^2 - \Delta^2 \right) }{e\,v_0\,v_z\,\Delta} \right] , \quad
\sigma^{({\rm conc})}_{zz}  = 
\frac{ -\,\tau\, e^4\, v_z^3\,k_0 \, v_0 \, \Delta^4 \, (B_x^2+2\,B_5^2)}
{16\, \pi\, \mu^7},
\nn
\end{align}
\begin{align}
\label{eqset2_ah}
\bar \sigma_{xz}  &= 0\,, \quad
 \sigma^{\text{(ah)}}_{yz}  = 
\frac{ -\, e^3\, v_z \, v_0 \, k_0 \,\Delta^2 \, B_x} {16 \, \pi\, \mu^4} 
 \left[
1 +\frac{ 9\, e^2  \,v_z^2 \, v_0^2  \, \Delta^2 \left(B_x^2 + 4\,B_5^2 \right) }
 { 4\, \mu^6 } -\frac{3\,B_5\,e\,v_0\,v_z\,\Delta} {\mu^3}  \right].
\end{align}
Here too, we note that the $\theta $-dependence of the longitudinal component, $\bar \sigma_{zz}$, arises solely from the terms containing $B_x^2$; $B_5$- and $B_5^2$-dependent terms merely shift the values in a $ B_x^2$-independent way. The planar-Hall component, $\bar \sigma_{xz}$, is identically zero.
For the $\check L$-induced parts, the various terms arising in the intermediate steps are shown in Appendix~\ref{appset3}.
The overall contributions to the in-plane components turn out to be
\begin{align}
\sigma^{\text{(lf)}}_{zz} & = \frac{ \tau^3\, e^4\,v_z^3\, v_0   }
{ 8 \,\pi\, \mu^4 } \left[
-2\,  B_5 \,(B_5^2 + 3\, B_x^2) \,e\, k_0\, v_0\, v_z \,\Delta
+(2\, B_5^2 + B_x^2)\, k_0 \,\mu\, (\Delta^2 - \mu^2)
+3 \,B_5 \,B_x^2\, e \,v_z\, \Delta \,\zeta \right], 
\nn \sigma^{\text{(lf)}}_{xz} & = \frac{\tau^3 \,  e^5\,   k_0 \, v_0^4 \, v_z^2 \, \Delta \, B_5 \,B_x \,B_z }
{ 4 \, \pi \, \mu^4\,  \zeta }
\left( k_0 \, v_0 - \zeta \right).
\end{align}
On adding up all the out-of-plane parts, we get
\begin{align}
\label{eqset2_lf_oop}
\sigma^{\text{(lf)}}_{yz} & = 
\frac{\tau^2 \, e^3 \,  v_z \,  v_0 \,B_x } {128 \,\pi\,\mu^8 \,\zeta }
\Bigg[ 
16  \, \zeta  \,  k_0 \,  \mu^6 (\Delta^2 -\mu^2 ) 
- 32  \, B_5  \, \Delta  \,  e \,  \zeta  \,  k_0 \,\mu^5  \, v_0 \,  v_z
-3  \, \Delta^2 \,  e^2  \, v_0^2 \,  v_z^2\, 
\zeta  \,  k_0 \left(B_x^2 + 4 \,  B_5^2\right) 
\left(4  \, \Delta^2-7  \, \mu^2\right)
\nn & \hspace{ 2.75 cm }
-6  \, \Delta^2 \,  e^2  \, v_0 \,  v_z^2   \, \mu^4  \, B_x^2
+ 8  \, \Delta^2 \,  e^2 \,  k_0  \, \mu^2  \, v_0^4  \, B_z^2 \left(k_0 \,  v_0-\zeta \right)   
+ 24 \,  e^2 \,  \zeta  \,  k_0^2 \,  \mu^4  \, \tau^2 \,  v_0^3 \,  B_x^2  \, v_z^2 
\left(k_0 \,  v_0-\zeta \right)
\nn & \hspace{ 2.75 cm }
+ 16 \,  e^2 \,  k_0 \,  \mu^4  \, \tau^2  \, v_0^4  \, B_z^2 
\left(\zeta -k_0 \,  v_0\right)^2 \left(\zeta +2  \, k_0 \,  v_0\right)
+ 48 \,  B_5^2  \, e^2 \,  \zeta  \,  k_0  \, \mu^4  \, \tau^2 \,  v_0^2  \, v_z^2
\left(\mu^2-\Delta^2\right) 
\Bigg]\,.
\end{align}
Fig.~\ref{figset3} demonstrates the characteristics of the conductivity curves for some representative values of the parameters.
For the longitudinal component, $\Delta \sigma_{zz}^{\rm(conc)}$ is of the same order of magnitude as $\Delta \sigma_{xx}^{\rm (lf)}$, which are smaller than $\Delta \sigma_{xx}^{\rm(bc)}$ and $\Delta \sigma_{xx}^{\rm(m)}$ (and, hence, $\Delta \bar \sigma_{xx}$) by orders of magnitude (precisely, smaller by $  \sim 10^{-4} $). 
The out-of-plane component's sub-parts, $\Delta \sigma_{yz}^{(\delta)}$, are smaller than $\Delta \bar \sigma_{xx}$ by order of $10^{-6}$.

\section{Discussion of the results}
\label{secdis}

In this section, we provide a detailed and comprehensive discussion of the characteristics of the various components of the conductivity tensors for the 3 set-ups, as conveyed by the analytical expressions as well as representative plots.
A summary of all the results for the 3 set-ups is provided in Table~\ref{taball}.
We have condensed the expressions by defining
\begin{align}
    \Upsilon_1  &=  \frac{\tau \, e^4 \, v_z \, v_0^3  \, \Delta^2 }
{128 \, \pi \, \mu^7} \left[
k_0 \left( \mu^2 - \Delta^2 \right) 
+
\left \lbrace - \, 3\, k_0 \left( 2 \, \mu^2 - \Delta^2 \right)
+ \frac{2 \,\mu^4}
{v_0\, \zeta}  
\right \rbrace \right ],\quad
\Upsilon_2  =  \frac{\tau \, e^4  \,v_z^3\, v_0 \,k_0 \, \Delta^2}
{ 32 \, \pi \, \mu^7} 
\left[  \left( \mu^2 - \Delta^2 \right)
-3\left(   2 \, \mu^2 - \Delta^2 \right)
\right ] ,\nn 
\Upsilon_3 & = \frac{ -\, e^3\, v_z \, v_0 \, k_0 \,\Delta^2 }
 {16 \, \pi\, \mu^4} \,,\quad
 \Upsilon_4  = \frac{ 9\, e^2  \,v_z^2 \, v_0^2  \, \Delta^2 }
 { 4\, \mu^6 }  \,.
\end{align}

\begin{table}[ht!]
\subfigure[Nonzero components of $\bar \sigma $ and $\sigma^{\rm{(ah)}} $ components across the three set-ups. $B$ represents magnitude of actual magnetic field in the respective set-up.]
{\centering 
\begin{tabular}{|l|c|c|c|}
\hline
& Longitudinal & In-plane transverse & Out-of-plane transverse \\
\hline
Set-up~I &
$\Upsilon_1(B_x^2 + 3\, B_y^2)+ B_5\text{-term}+ B_5^2 \text{-term}$ &
$-2\, \Upsilon_1\, B_x\, B_y $&
$\Upsilon_3\, B_y \left[1+ \Upsilon_4 ( B^2 + 4\, B_5^2) \right]+ B_5\text{-term}$ \\[4pt]\hline
Set-up~II &
$\Upsilon_1 \, B_x^2 + B_5 \text{-term}
 + B_5^2 \text{-term}$ &
$0$ &
$0$ \\[4pt]\hline
Set-up~III &
$\Upsilon_2 \, (B_x^2 + 2\, B_5^2)
 + B_5\text{-term}$ &
$0$ &
$\Upsilon_3\, B_x\left[ 1+ \Upsilon_4 \, ( B_x^2 + 4 \, B_5^2)\right]
 + B_5\text{-term}$ \\ \hline
\end{tabular}}
\subfigure[Set-up~I: $B_5$- and $B_i$-dependence in LF-induced terms, organised by the values of $n$ and physical origins.]
{\centering 
\begin{tabular}{|l|c|c|c|c|}
\hline
 & $\sigma^{\text{lf,(h)}}$ & $\sigma^{\text{lf,(bc)}}$ & $\sigma^{\text{lf,(m)}}$ & $\sigma^{\text{lf, (conc)} }$ \\
\hline
$n=1$, $\sigma_{zx}$ & $ B_y$ &
\makecell{$ B_y\,B_5 ,\, B_y \,B_5^2, \,B_y \,( B_x^2 + B_y^2)$} &
\makecell{$B_y \, B_5, B_y\, B_5^2, \, B_y \,(B_x^2 + B_y^2 )$} &
$  B_y  \, B_5^2, \, B_y \,(B_x^2 + B_y^2 )$ \\[4pt]\hline
$n=2$, $\sigma_{xx}$ &
$ B_5^2, \,B_x^2 ,\, B_y^2$ &
\makecell{$  B_5\,  B_x^2, \, B_5\, B_y^2 , \, B_5^3$ } &
$  B_5\,  B_x^2, \, B_5\, B_y^2, \, B_5^3 $ &
$0$ \\[4pt]\hline
$n=2$, $\sigma_{yx}$ &
$ B_y \, B_x$ &
$ B_y\,  B_x\, B_5 $ &
$ B_y\, B_x \,B_5 $ &
$0$ \\[4pt] \hline
$n=3$, $\sigma_{zx}$ &
\makecell{$ B_y \,B_5^2, \, B_y\, ( B_x^2 + B_y^2)$} &
$0$ & $0$ & $0$ \\ \hline
\end{tabular}}
\subfigure[Set-up~II: $B_5$- and $B_i$-dependence in LF-induced terms, organised by the values of $n$ and physical origins.]
{\centering 
\begin{tabular}{|l|c|c|c|c|}
\hline
 & $\sigma^{\text{lf,(h)}}$ & $\sigma^{\text{lf,(bc)}}$ & $\sigma^{\text{lf,(m)}}$ & $\sigma^{\text{lf, (conc)} }$ \\
\hline
$n=1$, $\sigma_{yx}$ & $0$ & $0$ &
$ B_z \, B_5,\;  B_z\,B_5^2 ,\; B_z \,B_x^2   $ & $ B_z \,B_5^2 $ \\[4pt]\hline
$n=2$, $\sigma_{xx}$ & $ B_z^2,\; B_5^2,\; B_x^2$ &
$B_5 \, B_x^2,\; B_5 \, B_z^2, \, B_5^3 $ &
$B_5 \, B_x^2,\; B_5 \, B_z^2, \, B_5^3 $& $0$ \\[4pt]\hline
$n=2$, $\sigma_{zx}$ & $0$ &
$  B_x\, B_z\, B_5$ & $  B_x \,B_z\, B_5 $ & $0$ \\[4pt] \hline
$n=3$, $\sigma_{yx}$ &
$B_z \, B_x^2, \, B_z\,B_5^2,\,B_z^3 $ & $0$ & $0$ & $0$ \\ \hline
\end{tabular}}
\subfigure[Set-up~III: $B_5$- and $B_i$-dependence in LF-induced terms, organised by the values of $n$ and physical origins.]
{\centering 
\begin{tabular}{|l|c|c|c|c|}
\hline & $\sigma^{\text{lf,(h)}}$ & $\sigma^{\text{lf,(bc)}}$ & $\sigma^{\text{lf,(m)}}$ & $\sigma^{\text{lf, (conc)} }$ \\
\hline
$n=1$, $\sigma_{yz}$ &
$ B_x$ &
$ B_x \, B_5,\; B_x \, B_5^2 ,\;  B_x^3$ &
$ B_x \, B_5,\; B_x \, B_5^2 ,\;  B_x^3$  &
$ B_x \, B_z^2,\; B_x\,B_5^2 ,\; B_x^3$ \\[4pt]\hline
$n=2$, $\sigma_{zz}$ &
$ B_5^2,\; B_x^2$ &
$ B_5 \, B_z^2,\; B_5\, B_x^2,\; B_5^3$ &
$ B_5 \, B_z^2,\; B_5 \, B_x^2,\; B_5^3 $ & $0$ \\[4pt]\hline
$n=2$, $\sigma_{xz}$ & $0$ &
$ B_x \, B_z\, B_5 $ & $ B_x \, B_z\,  B_5 $ & $0$ \\[4pt]\hline
$n=3$, $\sigma_{yz}$ &
$ B_x \, B_z^2,\;  B_x \, B_5^2,\; B_x^3$ & $0$ & $0$ & $0$ \\ \hline
\end{tabular}}
\caption{\label{taball}Comparison of the overall behaviour of all nonzero components of the magnetoelectric conductivity for the three set-ups, when the chemical potential cuts the $s=2$ band at $T=0$.}
\end{table}

From the expressions of conductivity across the 3 set-ups, it is worth contrasting the peculiarities with the results for isotropic nodal-point semimetals, harbouring BC-monopoles, considering the example of a Weyl node with chirality $\chi $.
There, a linear-in-$B_5$ term ($\bs B_5$ is a constant there with no angular dependence) can appear only from $ \propto \tau^2\, \sum_{i=x, y} c_i \,[ B_i + \chi \, (B_5)_i ]^2$~\cite{ips_rahul_ph_strain, ips-ruiz, ips-rsw-ph}, which is $\propto \chi \,B\, B_5$. There is no other linear-in-$B_5$ term which is independent of $B$, because the hedgehog structure of the BC and the OMM of the Weyl node ensures that any integrand containing a single power of the topological quantities gives vanishing result (on integrating over the spherical Fermi surface). In the GNR, the vortex-like flux lines of $\bs B_5$, BC, and OMM are co-aligned,
preventing this zero result. In a Weyl semimetal (WSM), the nodes appear in conjugate pairs of $\chi = \pm 1 $ --- thus a sum over the nodes gets rid of the term $\propto \chi\,B\, B_5$ as well, leaving no linear-in-$B_5$ or linear-in-$B$ term.

\subsection{Set-up~I}

\subsubsection{Longitudinal conductivity: \texorpdfstring{$ \bar \sigma_{xx}$}{σxxset1}}

In the absence of strain (i.e., $\bs{B_5}= \bs 0$), $\sigma^{\rm (bc)}_{xx}$ is positive and
$\propto (B_x^2 + 3\,B_y^2)$. $\sigma^{\rm (m)}_{xx}$ shows the same $\propto (B_x^2 + 3\,B_y^2)$-dependence, but with a
larger coefficient of opposite sign. $\sigma^{\rm (conc)}_{xx}$ is also negative.
Therefore, it is the OMM that dictates the sign of the total non-Drude longitudinal response.

In the presence of nonzero $\bs{B_5} $, new terms appear as described below [cf. Eq.~\eqref{eqsigxxset1}]:
\begin{enumerate}

\item \emph{Linear-in-$B_5$ terms:} 
Linear-in-$B_5$ terms, independent of $\bs B$, appear. They are nonvanishing because the vortex structure of $\bs{B_5}(\phi)$
and $\bs{\Omega}_s(\phi)$ produces a $\phi$-independent value in the integrand from
$\bs{B_5}\cdot\bs{\Omega}_s $, which prevents the integral from vanishing upon $\phi$-integration.
Here, the BC-only and OMM-only parts carry \emph{opposite} signs. Of course there cannot be any concurrent term $\propto B_5$, because concurrence involves at least two terms in the integrand arising from the topological quantities.

\item \emph{Quadratic-in-$B_5$ terms:} Quadratic-in-$B_5$ terms appear in all the three parts, viz. BC, OMM, and concurrent. The effective field combination shifts from $(B_x^2 + 3\,B_y^2)$
to $(B_x^2 + 3\,B_y^2 + 4\,B_5^2)$ for the BC-only part while the OMM-only parts gets two competing contributions, i.e., a negative contribution from $ \left[- \left(B_x^2 + 3 \, B_y^2 + 4\, B_5^2 \right)  k_0  \left( 6\, \mu^2 - 5\, \Delta^2 \right) \right]$ and a positive contribution weighted by $\frac{2 \left(B_x^2 + 3 \, B_y^2 \right) \mu^4}  {v_0\, \zeta}$. Overall, the BC-only part is opposite in sign to the OMM-only and concurrent contributions.

\end{enumerate}

\subsubsection{Planar-Hall conductivity:
\texorpdfstring{$\bar \sigma_{yx}$}{σyxset1}}

$\bar \sigma_{yx}$ is \emph{completely blind} to the pseudomagnetic field.
Every $B_5$-dependent integrand for this component depends on  $\phi$ in a way which yields zero upon $\phi$-integration.
As a result, $\bar \sigma_{yx}$ retains precisely the form
$\bar \sigma_{yx} \propto B_x\,B_y$ found in
Ref.~\cite{ips-nlsm-ph}. From an experimental point of view, $\bar \sigma_{yx}$ thus provides a
clean reference that is decoupled from the lattice
deformation and, hence, can be used to benchmark the intrinsic topological
response independently of the strain on the sample.

\subsubsection{LF-induced in-plane contributions: \texorpdfstring{$\sigma^{\text{(lf)}}_{xx}$}{σxxlfset1} and \texorpdfstring{$\sigma^{\text{(lf)}}_{yx}$}{σyxlfset1}}

For $\bs{B_5} = \bs 0$, the only in-plane LF-induced parts arise
from the $n = 2$ term in $\mathcal Y_s$~\cite{ips-nlsm-ph}.
Strain induces additional terms (cf. Appendix~\ref{appset1}): (I) From $\bs{\mathcal{ N }}_{2,2}$, a term
$\propto B_5^2 \,(\Delta^2 - \mu^2) < 0$ appears in
$\sigma^{\text{lf},(h)}_{xx}$.
(II) From $\bs{\mathcal{ N }}_{2,3}$, terms proportional to $B_5$ emerge in both
$\sigma^{\text{lf,(bc)}}_{xx}$ and
$\sigma^{\text{lf,(m)}}_{xx}$.
In $\sigma^{\text{lf,(bc)}}_{xx}$, these are proportional to $B_5^3$, $B_5\,B_x^2$, and $B_5\,B_y^2$, respectively.
For $\sigma^{\text{(lf)}}_{yx}$, it is $ \propto B_5\,B_x\,B_y$.

\subsubsection{Out-of-plane conductivity: \texorpdfstring{$\sigma_{zx}$}{σzxOPset1}}

The overall structures of the anomalous-Hall and LF-induced
response [cf. Eqs.~\eqref{eqsigyxset1} and \eqref{eqsigzsxset1}] are preserved with respect to the $\bs{B_5} = \bs 0$ results \cite{ips-nlsm-ph}: the $n=1,3$ terms of Eq.~\eqref{lf_cond1} produce out-of-plane response, while
$n=2$ produces in-plane (longitudinal and transverse) response (cf. Appendix~\ref{appset1}). On turning on $\bs{B_5}$, we observe the following:

\begin{enumerate}

\item \emph{Anomalous-Hall component:}
The only non-LF-sourced out-of-plane conductivity in set-up~I is the
anomalous-Hall component, $\sigma^{\text{(ah)}}_{zx}$, driven by the
anomalous velocity $ \propto \bs{E} \times \bs{\Omega}_s$
along $\hat{\bs{z}}$. For $\bs{B_5}= \bs 0$ this
response has 2 terms: one $\propto B_y$ and another $ \propto B_y\,(B_x^2 + 3\,B_y^2) $.
Strain introduces two new $B_5$-dependent terms --- a term $\propto B_y \, B_5$ with a positive coefficient, and a negative
quadratic shift of $4\, B_5^2$ in the $(B_x^2 + 3\,B_y^2)$ part.

\item \emph{LF-induced out-of-plane response:}
For $\bs{\mathcal{N }}_{1,2}$, new contributions linear in $B_5$ appear
in both $\sigma^{\text{lf,(bc)}}_{zx}$ and
$\sigma^{\text{lf,(m)}}_{zx}$, proportional to $B_5\,B_y$.
For $\bs{\mathcal{N }}_{1,3}$, the BC-only contribution acquires the clean
enhancement $B_y \,(B_x^2 + B_y^2) \to B_y\,(B_x^2 + B_y^2 + 4\,B_5^2)$. The OMM-only and concurrent terms also acquire
$\propto B_5^2\,B_y$ terms, but they do not reduce to a form $\propto B_y \, (B_x^2 + B_y^2 + 4\,B_5^2)$.
$\bs{\mathcal{ N }}_{3,3}$-generated term acquires
a $\propto B_y\,B_5^2 $ contribution.

\end{enumerate}

\subsection{Set-up~II}
\label{sec:disc_set2}
Here, the planar-Hall ($\bar \sigma_{zx}$) and the anomalous-Hall ($\sigma^{\text{(ah)}}_{yx} $) parts are identically zero.

\subsubsection{Longitudinal conductivity: \texorpdfstring{$ \bar \sigma_{xx}$}{σxxset2}}

For $\bs{B_5}=\bs 0$, the BC-only part is positive, the OMM-only part is negative and
dominant. The concurrent part is also negative, resulting in a net negative value for the non-Drude part of $\bar \sigma_{xx}$~\cite{ips-nlsm-ph}. Strain introduces the following extra terms [cf. Eq.~\eqref{eqset2long}]:
\begin{enumerate}

\item \emph{Linear-in-$B_5$ terms:}
Both $\sigma^{\rm (bc)}_{xx}$ and $\sigma^{\rm (m)}_{xx}$ acquire terms
linear-in-$B_5$.
The BC-only and the OMM-only coefficients of these linear-$B_5$ have opposite signs, similar to set-up I.

\item \emph{Quadratic-in-$B_5$ terms:} Terms $\propto B_5^2$ appear such that
there is a shift from $B_x^2$ to $(B_x^2 + 4\,B_5^2)$ in the terms not having a $\mu^4$-dependence.
Their coefficients in the BC-only term are opposite to those of OMM-only and concurrent parts.

\end{enumerate}

\subsubsection{LF-induced contributions: \texorpdfstring{$\sigma^{\text{(lf)}}_{xx}$}{σxxlfset2} and \texorpdfstring{$\sigma^{\text{(lf)}}_{zx}$}{σzxlfset2}}

For $\bs{B_5}= \bs 0$, the sole in-plane contribution is obtained for $n = 2$, such that
$\sigma^{\text{lf,(bc)}}_{xx} = \sigma^{\text{lf,(m)}}_{xx}
= \sigma^{\text{lf,(conc)}}_{xx} = 0$ --- only $\sigma^{\text{lf,(h)}}_{xx}$ is nonzero \cite{ips-nlsm-ph}.
Strain adds nonzero LF-induced contributions as follows (cf. Appendix~\ref{appset2}): (I) From $\bs{\mathcal{N }}_{2,2}$, a term
$\propto  B_5^2\, (\Delta^2 - \mu^2) < 0$ is added to
$\sigma^{\text{lf,(h)}}_{xx}$ in the same manner as in set-up~I.
(II) From $\bs{\mathcal{N }}_{2,3}$, nonzero $B_5$ makes LF-induced
corrections emerge in both $\sigma^{\text{(lf)}}_{xx}$ and $\sigma^{\text{(lf)}}_{zx}$.
In the former, these are proportional to $B_5\,B_x^2$, $B_5\,B_z^2$, and
$B_5^3$. $\sigma^{\text{(lf)}}_{zx}$ gets contributions $\propto B_5\,B_x\,B_z$.

\subsubsection{Out-of-plane conductivity: \texorpdfstring{$\sigma_{yx}$}{σyxOPset2}}

Set-up~II is the only configuration where the out-of-plane response is
entirely of LF origin (i.e., has no contribution from anomalous-Hall effect). Here, the $\bs{\mathcal{N }}_{1,1}$ part for the unstrained case is zero \cite{ips-nlsm-ph}, unlike set-ups~I and III. This makes it parametrically weaker than that in set-ups~I and~III. Overall, the $n=1,3$ terms of Eq.~\eqref{lf_cond1} produce out-of-plane response, while
$n=2$ produces in-plane (longitudinal and transverse) response (cf. Appendix~\ref{appset2}). Strain induces additional terms as follows:

\begin{enumerate}

\item \emph{Absence of anomalous-Hall part of the response:}
The anomalous-Hall component, $(\sigma^{\text{(ah)}})_{yx}$, vanishes identically as a consequence of $\Omega^z_s = 0$.

\item \emph{LF-induced out-of-plane response:}
For $\bs{\mathcal{N }}_{1,2}$, a nonzero OMM-only contribution $ \propto B_z \,B_5 $ appears.
For $\bs{\mathcal{N }}_{1,3}$, the OMM-only contribution acquires a
$\propto B_5^2\,B_z$ term,
and the concurrent part gains a $ \propto B_z \, B_5^2 $ term (being nonzero only in the presence of strain).
$\bs{\mathcal{N }}_{3,3}$ gains a $\propto B_z \,B_5^2$ contribution.

\end{enumerate}

\subsection{Set-up~III}
Here, the planar-Hall part, $\bar \sigma_{xz}$, is identically zero.

\subsubsection{Longitudinal conductivity:
\texorpdfstring{$\bar \sigma_{zz}$}{σzzset3}}

For $\bs{B_5}= \bs 0$, the non-Drude part of $\bar \sigma_{zz}$ has a positive BC-only part $ \propto B_x^2$, while $\sigma^{\rm (m)}_{zz}$
comes with an opposite sign with a larger magnitude.
Since $\sigma^{\rm (conc)}_{zz}$ is also negative, the net non-Drude
response is negative and dominated by the OMM, analogous to
set-ups~I and~II. $\bar\sigma_{xz}$ vanishes identically, irrespective of the presence/absence of strain.
Strain introduces extra terms as follows [cf. Eq.~\eqref{eqset3long}]:
\begin{enumerate}

\item \emph{Linear-in-$B_5$ terms:}
Both $\sigma^{\rm (bc)}_{zz}$ and $\sigma^{\rm (m)}_{zz}$ acquire terms linear-in-$B_5$.
The BC-only and the OMM-only coefficients are opposite in signs. Of course, there is no scope of getting a concurrent term here.

\item \emph{Quadratic-in-$B_5$ terms:}
An effective field combination of $(B_x^2 + 2\,B_5^2)$ is observed in all three non-Drude parts. The $B_5^2$ coefficient of~2 (rather than~4, which appears in the first 2 set-ups) is a direct consequence of $\bs{E} \parallel \bs{\hat{z}}$ --- only one of the two in-plane components of $\bs{B_5}$ contributes to
the integral, compared to both in-plane components contributing in the $xx$ response of set-ups~I
and~II. The BC-only contribution from $B_5^2$ is positive, while the coefficients of the OMM-only and the concurrent parts are negative --- the net $\propto B_5^2$ part is overall negative reflecting the dominance of the OMM-contributed parts.

\end{enumerate}

\subsubsection{LF-induced in-plane contributions: \texorpdfstring{$\sigma_{zz}^{\rm (lf)}$}{σzzset3lf}
and \texorpdfstring{$\sigma_{xz}^{\rm (lf)}$}{σxzset3lf}}

For $\bs{B_5}=\bs 0$, the sole in-plane LF-induced contributions appear from $\bs{\mathcal{ N }}_{2,2}$~\cite{ips-nlsm-ph}, being
$ \propto B_x^2   \left(\Delta^2 - \mu^2\right) < 0$,
with $\sigma^{\text{lf,(bc)}}_{zz} = \sigma^{\text{lf,(m)}}_{zz} = \sigma^{\text{lf,(conc)}}_{zz} = 0$.
Unlike set-up~II, there is no $B_z^2$ contribution from $\bs{\mathcal{ N }}_{2,2}$.
Strain introduces additional terms as follows (cf. Appendix~\ref{appset3}):
\begin{enumerate}

\item For $\bs{\mathcal{N }}_{2,2}$, $B_x^2$ is replaced by
$(B_x^2 + 2\,B_5^2)$, accompanying $ (\Delta^2 - \mu^2) $.

\item For $\bs{\mathcal{N }}_{2,3}$, linear-in-$B_5$ and cubic-in-$B_5$ terms emerge in $\sigma^{\text{lf,(bc)}}_{zz}$ and
$\sigma^{\text{lf,(m)}}_{zz}$.
These are proportional to $B_5\,B_x^2$, $B_5\,B_z^2$, and $B_5^3$. Additionally, $\bs{\mathcal{ N }}_{2,3}$ generates
contributions  in $\sigma^{\text{lf,(bc)}}_{xz}$ and
$\sigma^{\text{lf,(m)}}_{xz}$, which are $ \propto B_5\,B_x\,B_z$.

\end{enumerate}


\subsubsection{Out-of-plane conductivity: \texorpdfstring{$\sigma_{yz}$}{σyzOPset3}}

The anomalous-Hall and LF-induced response [cf. Eqs.~\eqref{eqset2_ah} and \eqref{eqset2_lf_oop}] are preserved with respect to the $\bs{B_5} = \bs 0$ results \cite{ips-nlsm-ph}: the $n=1,3$ terms of Eq.~\eqref{lf_cond1} produce out-of-plane response, while
$n=2$ produces in-plane (longitudinal and transverse) response (cf. Appendix~\ref{appset3}). On turning on $\bs{B_5}$, we observe the following:
\begin{enumerate}

\item \emph{Anomalous-Hall component:} The $B_5$-dependent terms take the same structural form as
in set-up~I: a linear-in-$B_5$ with a negative coefficient and a quadratic-in-$B_5$ term, replacing $B_x^3 $ by  $B_x\, (B_x^2 + 4\,B_5^2)$.
The coefficient~4 in $4\, B_5^2$ reflects that both the components of $\bs{B_5}$ contribute.

\item \emph{LF-induced out-of-plane response:}
This one closely mirrors that of set-up~I, with $B_x$ playing the role of $B_y$ (in set-up~I). From $\bs{\mathcal{N }}_{1,2}$, terms in
$\sigma^{\text{lf,(bc)}}_{yz}$ and
$\sigma^{\text{lf,(m)}}_{yz}$ are generated which are $ \propto B_x \,B_5 $.
For $\bs{\mathcal{N }}_{1,3}$, the BC-only response acquires the clean
enhancement $B_x^2 \to (B_x^2 + 4\,B_5^2)$, with the same
coefficient~4 as in set-up~I.
The OMM-only and concurrent contributions from $\bs{\mathcal{N }}_{1,3}$ also
acquire $\propto B_x \,B_5^2$ terms.
$\bs{\mathcal{ N }}_{3,3}$ generates a term $ \propto B_x\,B_5^2$.

\end{enumerate}

\section{Summary, concluding remarks, and future perspectives}
\label{secsum}

In our earlier work \cite{ips-nlsm-ph}, we investigated the nature of linear magnetoconductivity arising from the 3 configurations shown in Fig.~\ref{figsetup}(b). In this paper, we have supplemented the external electromagnetic fields of those set-ups with an intrinsic strain-induced axial pseudomagnetic field, $\bs B_5$. The main aim is to figure out if the chiral nature of $\bs B_5$ helps to succinctly portray the characteristics of the underlying BC and OMM, which also possess the same chiral structure of the vector-field lines.
In other words, the central question motivating this work is whether the antipodal
coupling structure of a strain-induced axial pseudomagnetic field can leave qualitatively distinct imprints on the magnetoelectric conductivity of a GNR.  The answer is unambiguously affirmative and it rests on
using a $\bs{B_5}(\phi)$ with the same vortex-like field lines as the BC and the OMM. This leads to the most significant feature: the dot product of $\bs{B_5}$ with the BC or the OMM is proportional to $ B_5\,( \sin^2\phi+\cos^2\phi) = B_5$, which is angle-independent and integrates to nonzero constants over the Fermi surface. The analogous products for a uniform physical field $\bs{B}$
generate oscillatory $\sin\phi$ and $\cos\phi$ factors that
vanish upon $\phi$-integration. One $\bs{B_5}\cdot\bs{\Omega}_s $ factor enters
through $ D_s = [1+e\bs{B}^{\rm tot}\cdot\bs{\Omega}_s]^{-1}$, which
produces $B_5$ and $B_5^2$ contributions that do not cancel upon
$\phi$-integration --- this is one of the ways for $B_5 $- and $B_5^2$-dependent terms to originate.
The second source is the OMM-induced correction to the dispersion, viz. $\varepsilon^{\rm (m)}=-\bs{B}^{\rm tot}\cdot\boldsymbol{m}$. This source also leads to both linear-in-$B_5$ and quadratic-in-$B_5$ contributions. It is important to note that $\propto B_5$ terms are \textit{not} of the form of $B\,B_5 $, because they originate from a single power of either BC or an OMM-induced quantity.
The non-LF planar-Hall conductivity, $\bar \sigma_{yx}$, in set-up~I is completely
immune to $\bs{B_5}$ at all orders. This feature makes $\bar\sigma_{yx}$ a strain-insensitive internal
reference --- in a sample where both $\bs{B}$ and $\bs{B_5}$ are present, it isolates the topological response to the physical field without contamination from any lattice deformation. Thus, this robustness is experimentally invaluable because it
provides a reference that tracks only the topological response to the physical field, uncorrupted by strain, and can be used to
calibrate the pristine BC- and OMM-contributions independently.

The strain-insensitive planar-Hall channel identified in set-up~I is, in principle, distinguishable from the strain-sensitive channels of set-ups~II and~III by comparing devices under controlled uniaxial versus torsional strain, along the lines of the pseudomagnetic-field measurements already reported in a doped type-II Weyl semimetal~\cite{exp_gauge} and in an inhomogeneous acoustic crystal~\cite{nlr-acoustic}. CuTeO$_3$~\cite{cuteo_nlsm} and Co$_3$Sn$_2$S$_2$~\cite{enke}, both identified in the literature as good candidates for an ideal GNR, are natural targets for such measurements, given the near-flatness of their nodal rings close to the Fermi level. We stress that no direct experimental test of the linear-in-$B_5$ signature predicted here exists yet. We frame it as a concrete prediction of the present work, rather than as a comparison against existing data.

Strain-induced transport in nodal-point semimetals --- WSMs,
Dirac semimetals (DSMs), and semimetals carrying higher-than-1/2 pseudospin values~\cite{pikulin_gauge, grushin-strain-wsm, landsteiner_gauge, liu_gauge, ips_rahul_ph_strain, ips-ruiz, ips-rsw-ph, arjona18_rotational} --- has been studied extensively over the past decade. In both nodal-point and nodal-ring systems, a spatially nonuniform lattice deformation enters the low-energy Hamiltonian as an effective axial gauge potential. This is because strain modifies the inter-site hopping integrals, such that the resulting perturbation can be absorbed into a $\bs{k}$-space shift, that looks like a chirality-odd vector potential~\cite{pikulin_gauge, landsteiner_gauge, liu_gauge,
grushin-strain-wsm, arjona18_rotational, nlsm-strain, nlr-acoustic}. The transport consequences in the two cases are, however, qualitatively different as discussed below.
Let us explain the dissimilarities by considering isotropic WSMs, which harbour pairs of nodal points in the BZ --- each pair comprising opposite chiralities, $\chi=\pm1$. A torsional deformation of a nanowire, for example, can generate a spatially uniform axial field, $\bs{B_5}=B_5\, \hat{\bs{z}}$ parallel to the wire's axis~\cite{pikulin_gauge}, which couples to a node with amplitude $\chi \,B_5$.
Around each Weyl node, the BC ($\bs{\Omega}_s^{(\chi)}$) and the OMM ($\bs{m}^{(\chi)}$) assume a hedgehog form, $ \propto\chi\,\hat{\bs{k}}/k^2$.
The key quantity entering the phase-space factor, $D_s^{(\chi)}$, is the
projection $\bs{B_5}\cdot \bs{\Omega}_s^{(\chi)}$, which at the
$\chi$-node is proportional to $\cos\theta_{\bs{k}}/k^2$, where
$\theta_{\bs{k}}$ is the polar angle of a point on the Fermi surface.
Upon integrating over $\theta_{\bs{k}}$,
$\int_0^\pi d\theta_{\bs{k}} \cos\theta_{\bs{k}} \sin\theta_{\bs{k}}= 0 $ --- a consequence of the spherical symmetry of the node's dispersion. This null result follows from the
mismatch between the isotropic hedgehog structure of the BC and the fixed direction of the uniform axial field.
The same angular integration eliminates terms $\propto \bs{B_5}\cdot \boldsymbol{m}^{(\chi)}$ for the OMM-induced parts.
Consequently, the leading strain-induced effect in all
nodal-point semimetals is necessarily $ \propto \tau^2\, (\bs B + \chi \,\bs B_5)^2$~\cite{ips_rahul_ph_strain, ips-ruiz, ips-rsw-ph}. When summed over the nodes with $\chi = \pm 1$, the linear-in-$B_5$ (also, this is linear-in-$B$) term vanishes. An analogous effect is seen in the GNR, when $\bs{B_5} = B_5 \left(-\sin\phi\,\hat{\bs{x}}
+\cos\phi\,\hat{\bs{y}}\right)$ --- the term $ \propto \tau^2\, B\, B_5$ vanishes upon $\phi $-integration. Essentially, the ungapped system comprises nodes at $\phi$ and $\phi + \pi$ with opposite chiralities, which can be visualised as a pair of Weyl nodes with $\chi = \pm 1$. 

An important feature of the pseudomagnetic field is that, unlike the band-structure parameters, such as the gap or the Fermi velocity, the strain amplitude $B_5$ can be continuously varied in a single sample --- for instance by uniaxial pressure or by driving acoustic waves~\cite{pikulin_gauge, nlsm-strain, nlr-acoustic, exp_gauge}. Because various components contain
terms $\propto B_5$, reversing the direction of the deformation will provide the antisymmetric combination,
$[\sigma_{ij}(\bs B_5)-\sigma_{ij}(- \bs B_5)]/2$, which can serve to extract the linear-in-$B_5$ response.
The simultaneous measurement of the planar-Hall component ($\bar\sigma_{yx}$) of set-up I, which is totally strain-immune, serves as an intrinsic reference that distinguishes the topological response uncoupled with impurities or deformation. The strain immunity of $\bar\sigma_{yx}$ is a robust $T$-independent statement, rooted in the vortex-versus-planar angular mismatch between $\bs B_5(\phi)$/BC/OMM and the physical field $\bs B$, rather than an artifact of the zero-temperature limit. We further note that this robustness is directly relevant to experiments. Transport measurements on candidate GNR and strain-engineered platforms are typically performed in the cryogenic, $T\ll\mu$ regime (see, e.g., the low-temperature transport/ARPES characterisation of the near-ideal nodal-loop material CuTeO$_3$~\cite{cuteo_nlsm}, and the low-temperature realisation of strain- and inhomogeneity-induced flat pseudo-Landau levels in acoustic and Weyl/Dirac platforms~\cite{nlr-acoustic, grushin-strain-wsm}). In this regime, for the positive-energy band, a Sommerfeld expansion of the thermal-convolution formula [cf. Eq.~\eqref{eqsigmat}] yields:
$$\sigma_{ij}(T) = \sigma_{ij}(T=0) + \frac{\pi^2}{6}\left(\beta \, \mu\right)^{-2} \mu^2
\left.\frac{\partial^2 \sigma_{ij}(\varepsilon)}{\partial\varepsilon^2}\right|_{\varepsilon=\mu}
 + \mathcal{O}\!\left(\left(\beta \, \mu\right)^{-4} \right)\,.$$
In fact, the $B_5$-independence of $\bar\sigma_{yx}$ survives at any
nonzero low temperature. Since
$\sigma_{yx}(T\!=\!0,\varepsilon)$ carries no $B_5$-dependence for
any value of the chemical-potential argument $\varepsilon$, and the thermal convolution of Eq.~\eqref{eqsigmat} involves only this $B_5$-independent function and a $B_5$-independent kernel, the full
finite-$T$ result $\bar\sigma_{yx}(T)$ is $B_5$-independent for a nonzero $T$. The Sommerfeld expansion shown above only aids to estimate the size of the leading thermal correction at the low temperatures of
experimental relevance, and plays no role in the immunity itself. Thus, the strain-insensitivity of $\bar\sigma_{yx}$ is not an artifact of truncating the expansion at leading order --- the leading (low-temperature) terms we report are simply the ones of direct quantitative relevance to experiments.

It is worth reminding the reader regarding when the framework used here breaks down. The semiclassical Boltzmann treatment is
valid only in the weak-field limit,
$|\bs{B}^{\rm tot}|\ll\mu^2/(e \, v_0^2)$, where many Landau levels are occupied and the inter-Landau-level
spacing becomes negligible such that the dispersion remains effectively
continuous. Beyond this regime, the discretisation of the dispersion from the quantised Landau levels must be accounted for \cite{phe_nlsm}, where techniques like Kubo formalism must be used.  A related limitation is the RTA employed in our calculations. In a more realistic scenario, an exact treatment of the collision integrals \cite{timm, ips-exact-spin1, ips-exact-kwn, ips-exact-rsw, ips-dipole-vnr}, depending on the dominant channel of scattering processes, would help unravel the quantitative signatures. However, the computations of all possible components of conductivity in all the 3 planar-Hall set-ups warrant challenging numerical computations, which are beyond the scope of this work.

 The limitations of the RTA using a momentum-independent $\tau$, used throughout this work, is that it strips $\tau$
off any angular dependence that an anisotropic scattering
mechanism would carry. Let us illustrate it by an example. The identity,
$\bs\Omega_s(\kappa\!=\!0,\phi)=- \,\bs\Omega_s(\kappa\!=\!0,\phi+\pi)$, together with the corresponding relations for $\boldsymbol m(\bs k)$, $\boldsymbol v^{(0,s)}_\perp(\bs k)$, and
$\bs B_5(\phi)$, played the quintessential role to ensure that the $B\,B_5$ cross-term vanishes upon $\phi$-integration. Now if we promote $\tau\to\tau(\phi,\gamma)$, respecting $\tau(\phi+\pi,\gamma)=\tau(\phi,\gamma)$, then $B\,B_5$ cross-term
continues to be zero. However, this condition is not sufficient to guarantee the
$B_5$-independence of $\bar\sigma_{yx}$ found here: a $\tau$
that respects $\phi\to\phi+\pi$, but is anisotropic at higher angular order
[e.g., $\propto\cos (2\phi) $] can still generate a purely $B_5$-linear contribution through terms such as
$v_x^{(0,s)}v_y^{(0,s)}(\bs B_5\cdot\bs\Omega_s)$, whose intrinsic angular
structure [$\propto\sin (2\phi) $] couples directly to the corresponding
harmonic of $\tau$ [if we imagine writing $\tau(\phi,\gamma)$ as a Fourier series in $\phi$]. A genuinely momentum-dependent $\tau(\bs k)$ therefore cannot be assessed by symmetry arguments alone. In general, the proper angular dependence of $\tau$ automatically appears in the linearised Boltzmann equation in an exact formulation --- this is the one incorporating the
dominant scattering channel into the collision integral and solving the resulting integro-differential equation exactly along the lines of Refs.~\cite{timm, ips-exact-spin1, ips-exact-kwn, ips-exact-rsw}. Carrying out such a treatment for all three planar-Hall set-ups of a strained GNR is
a substantial undertaking in its own right and is beyond the scope of the
present work, as already stated earlier. Indeed, the failure of an RTA
has been observed in diverse contexts (see, for example,
Refs.~\cite{timm, ips-exact-spin1, ips-exact-kwn, ips-exact-rsw, rta1, rta2}),
where the correct physics is only captured by an exact treatment of the
collision integrals.

Several directions for future work present themselves naturally.  The
most immediate is the thermal and thermoelectric response of strained
GNRs. The OMM-distortion of the dispersion deserves a closer
attention. The calculation presented here is valid for low values of $ | \bs B^{\rm tot} |$,
which are below the threshold for the Lifshitz transition when the Fermi surface pinches off from a ring-cyclide to a horn-cyclide Fermi~\cite{yang1,yang_review_nlsm}. Computing the strain-induced conductivity at and after this transition will be an interesting direction to follow. Lastly, we note that a different type of nodal-ring semimetal is possible, dubbed as the vortex nodal-ring (VNR) \cite{vortex-nrsm, ips-dipole-vnr}, which does not have any $\mathcal P \mathcal T$-symmetry to start with and which has a nonzero BC-profile along all three momentum-directions. The nomenclature is related to the fact that the pseudospin vector-field [i.e., the vectorial function $\boldsymbol d (\bf k)$ in a two-band Hamiltonian of the form $\boldsymbol d (\bf k) \cdot \bs \sigma $] takes a vortex-like or smoke-ring texture in the momentum space. In the future, it will be worthwhile to work out the effects of pseudomagnetic fields in such systems.

\appendix

\section{Semiclassical Boltzmann-transport formalism}
\label{appboltz}

In this appendix, we review the semiclassical 
Boltzmann-transport formalism employed in the main text to extract the 
magnetoelectric conductivity of the GNR, following the derivations outlined in the literature~\cite{ips-kush-review, ips_rahul_ph_strain, rahul-jpcm, ips-ruiz, ips-rsw-ph, ips-spin1-ph}. Here we will assume that the net magnetic field is $\bs B$. The inclusion of strain for GNR is straightforward --- we just need to replace $\bs B$ by $( \bs B + \bs B_5 ) $.

For a Bloch wavepacket built out of the cell-periodic state of band $s$ and centred at 
$(\bs{r}, \bs{k})$, the semiclassical equations of motion are captured by 
\begin{align}
\label{eqEOMcoupled}
\dot{\bs{r}} 
= \bs{v}_s - \dot{\bs{k}} \cross \bs{\Omega}_s(\bs{k}) \, , 
\qquad
\dot{\bs{k}} = -\, e\, \bs{E} - e\, \dot{\bs{r}} \cross \bs{B} \, . 
\end{align} 
The wavepacket's energy comprises the bare band-dispersion modified by the OMM-induced Zeeman-like shift:
\begin{align}
\label{eqEeff}
\mathcal{E}_s(\bs{r}, \bs{k}) 
= \varepsilon_s(\bs{k})
+ \varepsilon^{\rm (m)}(\bs{k})\, , \qquad 
\varepsilon^{\rm (m)}(\bs{k}) 
= -\, \bs{m}_s(\bs{k}) \cdot \bs{B}\, . 
\end{align}
The OMM-corrected effective group velocity is
\begin{align}
\label{eqveff}
\bs{v}_s(\bs{k}) = \nabla_{\bs{k}}\mathcal{E}_s(\bs{k}) 
= \bs{v}^{(0,s)}(\bs{k}) + \bs{v}^{\rm (m)}(\bs{k}) \, , 
\qquad
\bs{v}^{(0,s)} = \nabla_{\bs{k}}\varepsilon_s\, , \qquad
\bs{v}^{\rm (m)} = \nabla_{\bs{k}}\varepsilon^{\rm (m)}\, . 
\end{align}
Here the pseudomagnetic contribution $\bs{B}_5$ enters on the same 
kinematic footing as the physical magnetic field
\cite{nlsm-strain, pikulin-gauge-nlsm, ips-rsw-ph,ips_rahul_ph_strain}. Solving 
Eq.~\eqref{eqEOMcoupled} for $\dot{\bs r}$ and $\dot{\bs k}$ separately 
gives the explicit forms
\begin{align}
\label{eqEOM}
\dot{\bs{r}} & = D_s(\bs{k}) 
\left[\, \bs{v}_s + e\, \bs{E} \cross \bs{\Omega}_s 
+ e\, (\bs{v}_s \cdot \bs{\Omega}_s)\, \bs{B} \,\right], \quad
\dot{\bs{k}} = -\, e\, D_s(\bs{k}) 
\left[\, \bs{E} + \bs{v}_s \cross \bs{B} 
+ e\, (\bs{E}\cdot \bs{B})\, \bs{\Omega}_s \,\right].
\end{align}

Kinetic theory replaces the trajectories of the quasiparticles by a phase-space distribution 
$f_s(\bs{r}, \bs{k}, t)$, whose evolution obeys the Boltzmann equation 
\cite{xiao_review, timm, ips-kush-review, ips-rsw-ph,ips-kush}
\begin{align}
\label{eqBE}
\partial_t f_s + \dot{\bs{r}}\cdot \nabla_{\bs{r}} f_s + \dot{\bs{k}}\cdot \nabla_{\bs{k}} f_s 
= \mathcal{I}_{\rm coll}[f_s]\, ,
\end{align}
where the drift terms on the left-hand side describe the deterministic 
Hamiltonian flow and the collision integral 
$\mathcal{I}_{\rm coll}[f_s]$ on the right encodes the stochastic effect 
of scattering. Substituting the solutions for $\dot{\bs r}$ and $ \dot{\bs k}$ from 
Eq.~\eqref{eqEOM} into Eq.~\eqref{eqBE}, we obtain the semiclassical 
Boltzmann equation (SBE):
\begin{align}
\label{eqSBE}
& D_s \Big[\, \left\{ \bs{v}_s 
+ e\, \bs{E}\cross \bs{\Omega}_s 
+ e\, (\bs{v}_s\cdot \bs{\Omega}_s)\, \bs{B} \right\} \cdot \nabla_{\bs{r}} 
-\, e\, \left(\bs{E} + \bs{v}_s \cross \bs{B}\right) 
\cdot \nabla_{\bs{k}} 
-\, e^2\, (\bs{E}\cdot \bs{B})\, \bs{\Omega}_s 
\cdot \nabla_{\bs{k}} 
\,\Big] f_s 
+ \partial_t f_s 
= \mathcal{I}_{\rm coll}[f_s] \, .
\end{align}
For the transport configurations considered in the main text, the 
external fields are spatially uniform and time-independent, so that 
$\nabla_{\bs r} f_s = 0$ and $\partial_t f_s = 0$. Now we resort to the  relaxation-time approximation (RTA), using
\begin{align}
\label{eqRTA}
\mathcal{I}_{\rm coll}[f_s(\bs{k})]
\sim -\, \frac{f_s(\bs k) - f_0 (\mathcal{E}_s(\bs{k}))} {\tau} \, .
\end{align}
Here, $\tau$ is the relaxation time, taken here as a phenomenological momentum-independent constant.

To extract the linear-response conductivity, we treat $\bs{E}$ as a 
small perturbation and parametrise the nonequilibrium distribution as
\begin{align}
\label{eqfsplit}
f_s(\bs{k}) = f_0(\mathcal{E}_s(\bs{k})) + \delta f_s(\bs{k})\, , 
\qquad 
\delta f_s(\bs{k}) 
= \left[-\, \partial_{\mathcal{E}_s} f_0(\mathcal{E}_s) \right]
\tilde{g}_s(\bs{k}) \, .
\end{align}
Substituting this into the SBE and retaining terms proportional to linear order in $|\bs{E}|$, one obtains
the linearised SBE, which takes the form of
\begin{align}
\label{eqLinSBE}
\left(1 - e\, \tau\, D_s\, \check{L}\right) \tilde{g}_s(\bs{k}) 
= -\, e\, \tau\, D_s\, \bs{E} \cdot \left (\bs{v}_s + \bs{u}_s\right ) .
\end{align}
The formal solution can be expressed as a Neumann-series, 
\begin{align}
\label{eqNeumann}
\tilde{g}_s(\bs{k}) 
= -\, e\, \tau\, \left[ D_s\, \bs{E}\cdot(\bs{v}_s + \bs{u}_s) 
+ \mathcal{Y}_s(\bs{k})\, \right] , 
\text{ where }
\mathcal{Y}_s(\bs{k}) = \sum_{n=1}^{\infty} \left(e\, \tau\, D_s\right)^n \check{L}^n 
\left[\, D_s\, \bs{E}\cdot (\bs{v}_s + \bs{u}_s) \,\right] . 
\end{align}

Here, $\check{L} = \left ({\boldsymbol  v}_s \times \bs{B} \right ) \cdot \nabla_{\bs{k}} $ is the Lorentz-force (LF) operator. Inserting the solution for $ \delta f_s$, the current density contributed by band $s$ decomposes cleanly into three 
physically distinct pieces: $\bs{J}_s = \bs{J}_s^{\text{(ah)}} + \bar{\bs{J}}_s + \bs{J}_s^{\text{lf}} $, where
\begin{align}
\label{eqJAH}
\bs{J}_s^{\text{(ah)}} 
= -\, e^2 \int \frac{d^3 \bs{k}}{(2\pi)^3}\, 
\left(\bs{E}\cross \bs{\Omega}_s\right) f_0(\mathcal{E}_s) \Rightarrow
\left(\sigma_s^{\text{(ah)}}\right)_{ij}
= -\, e^2\, \epsilon_{ijl} 
\int \frac{d^3 \bs{k}}{(2\pi)^3}\, \Omega_s^l\, 
f_0(\mathcal{E}_s) \, ;
\end{align}
\begin{align}
\label{eqJnonah}
& \bar{\bs{J}}_s 
= -\, e^2\, \tau \int \frac{d^3 \bs{k}}{(2\pi)^3}\, 
D_s\, \left[\bs{v}_s + \bs{u}_s\right] 
\left\{\left(\bs{v}_s + \bs{u}_s\right)\cdot \bs{E}\right\} 
\frac{\partial f_0(\mathcal{E}_s)}{\partial\mathcal{E}_s}
\nn &  \Rightarrow
\left(\bar\sigma_s\right)_{ij} 
= -\, e^2\, \tau \int \frac{d^3 \bs{k}}{(2\pi)^3}\, 
D_s\, \left[(\bs{v}_s)_i + (\bs{u}_s)_i\right] 
\left[(\bs{v}_s)_j + (\bs{u}_s)_j\right] 
\frac{\partial f_0(\mathcal{E}_s)}{\partial\mathcal{E}_s} \, ; 
\end{align}
and
\begin{align}
\label{eqJlf}
& \bs{J}_s^{\text{lf}} 
= -\, e^2\, \tau \int \frac{d^3 \bs{k}}{(2\pi)^3}\, 
\left[\bs{v}_s + \bs{u}_s\right] 
\, \frac{\partial f_0(\mathcal{E}_s)}{\partial\mathcal{E}_s}\, 
\mathcal{Y}_s(\bs{k}) \Rightarrow
\left(\sigma_s^{\text{(lf)}}\right)_{ij} 
= -\, e^2\, \tau \int \frac{d^3 \bs{k}}{(2\pi)^3}\, 
\left[(\bs{v}_s)_i + (\bs{u}_s)_i\right] 
\frac{\partial f_0(\mathcal{E}_s)}{\partial\mathcal{E}_s}\, 
\frac{\partial \mathcal{Y}_s}{\partial E_j} \, . 
\end{align}
Because each action of $\check L$ contributes one linear power of 
$|\bs{B}|$, and further powers arrive through $\bs{v}^{\rm (m)}$ 
and the expansions of $D_s$ and $ \partial_{\mathcal{E}_s} f_0 (\mathcal{E}_s) $, 
the individual terms of $\mathcal{Y}_s$ carry a nested structure in 
$|\bs{B}|$. Following the convention of 
Refs.~\cite{ips-tilted, ips-rsw-ph, ips-spin1-ph, ips-nlsm-ph}, we 
label each sub-contribution to the integrand of Eq.~\eqref{eqJlf} by 
$\bs{\mathcal{N}}_{n,p}$,
where $n$ counts the power of $\check L$ and $p \ge n$ is the total power of 
$|\bs{B}|$ in that piece. Truncating at cubic order in 
$|\bs{B}|$ retains contributions with $(n,p) \in \{(1,1), 
(1,2), (1,3), (2,2), (2,3), (3,3)\}$. Explicit expression for each 
$\bs{\mathcal{N}}_{n,p}$ and its evaluation in the three set-ups are 
demonstrated in Appendix~\ref{appLF}.

\section{Full expressions for the OMM-induced corrections to band-velocities}
\label{appvel}

In set-up I, dealt in Sec.~\ref{secset1}, we have $\bs{E} = E_x\, {\bs{\hat x}}$ and $ \bs{B}=B_x \,{\bs{\hat x}} + B_y \,{\bs{\hat y}} $. Consequently, Eq.~\eqref{eqtopo} translates into  $ \varepsilon^{\rm (m)} (\bs{k}) 
 = \frac{e\,v_z\, v_0\, \Delta} {2\, \epsilon^2}\, 
\left[ \frac{\left( B_x-B_5 \sin{\phi} \right)\, k_y - \left( B_y+B_5 \cos {\phi} \right)\, k_x } {k_{\perp}} \right]$. The full expression for the components of OMM-induced velocity-correction are as follows:
\begin{align}
 v^{\rm (m)}_x  & = \frac{ - \, e\,  v_0\, v_z \, \Delta}{2\, \epsilon^4\, k_{\perp}^3} 
\big[ 2\, v_{0}^2\, k_x \left(k_{\perp}^2 - k_{0}\, k_{\perp} \right)
\lbrace\left( B_x-B_5 \sin{\phi} \right)\, k_y - \left( B_y+B_5 \cos {\phi} \right)\, k_x \rbrace
\nn & \quad 
+ \epsilon^2\, k_y
\lbrace\left( B_x-B_5 \sin{\phi}  \right)\, k_x + \left( B_y+B_5 \cos {\phi} \right)\, k_y \rbrace \big]
\nn & = 
\frac{e \,  v_0\, v_z \, \Delta}{2 \, \epsilon^4} 
\big[
2 \, v_0^2 \, \kappa \cos{\gamma}
\cos{\phi} 
\lbrace   \left( B_y+B_5 \cos {\phi} \right) \cos {\phi} - \left( B_x-B_5 \sin{\phi}  \right)  \sin{\phi} \rbrace
\nn & \qquad - \frac{ \epsilon^2 \sin{\phi}
\lbrace \left( B_x-B_5\sin{\phi}  \right) \cos{\phi}  + \left( B_y+B_5 \cos {\phi} \right) \sin {\phi} \rbrace }
{k_0 + \kappa \cos \gamma}  
\big],
 \nn  v^{\rm (m)}_y & = 
 \frac{ -\, e \, v_0\, v_z \, \Delta}{2\, \epsilon^4\, k_\perp^3} 
\big[ 2\, v_0^2\, k_y \left(k_\perp^2 - k_0\,k_\perp\right)\, 
\lbrace \left( B_x-B_5 \sin{\phi}  \right)\, k_y - \left( B_y+B_5 \cos {\phi} \right)\, k_x \rbrace
\nn & \quad  - \epsilon^2\, k_x
 \lbrace \left( B_x-B_5 \sin{\phi}  \right)\, k_x + \left( B_y+B_5 \cos {\phi} \right)\, k_y \rbrace 
\big]\nn
& = 
\frac{ e \, v_0\, v_z \, \Delta}{2 \, \epsilon^4} 
\big[ 2 \,v_0^2 \, \kappa \cos {\gamma }
\sin {\phi}
\lbrace  \left( B_y+B_5 \cos {\phi} \right) \cos{\phi}  - \left( B_x-B_5 \sin{\phi}  \right) \sin {\phi} \rbrace 
\nn & \quad + \frac{ 
\epsilon^2 \cos{\phi}
\lbrace \left( B_y+B_5 \cos {\phi} \right)  \sin{\phi} +\left( B_x-B_5 \sin{\phi}  \right) \cos {\phi} \rbrace }
{k_0 + \kappa \cos {\gamma}}  
\big],
 \nn v^{\rm (m)}_z  & = 
\frac{ e \, v_0\, v_z^3 \, \Delta \,k_z }
{\epsilon^4\, k_\perp} 
\left[ \left( B_y+B_5 \cos {\phi} \right)\,k_x - \left( B_x-B_5 \sin{\phi} \right)\,k_y \right]
\nn & =
\frac{e  \, v_0^2 \, v_z^2 \, \Delta \, \kappa \sin {\gamma} } {\epsilon^4} 
\left[
\left( B_y+B_5 \cos {\phi} \right) \, \cos{\phi} - \left( B_x-B_5\sin{\phi}  \right)  \sin{\phi} \right ].
\end{align}

For set-up II, considered in Sec.~\ref{secset2}, we have $ \varepsilon^{\rm (m)} (\bs{k}) 
  = \frac{e\,v_z\, v_0\, \Delta}
{2\, \epsilon^2}
\left[ \frac{\left( B_x-B_5 \sin{\phi} \right)\, k_y - B_5 \cos {\phi} \, k_x } {k_{\perp}} 
 \right]$. This results in
\begin{align}
v^{\rm (m)}_x & =
 \frac{ - e\,  v_0\, v_z \, \Delta}{2\, \epsilon^4\, k_{\perp}^3} 
\left[ 2\, v_{0}^2\, k_x \left(k_{\perp}^2 - k_{0}\, k_{\perp} \right)
\lbrace
\left( B_x-B_5 \sin{\phi} \right)\, k_y - B_5 \cos {\phi} \, k_x \rbrace
+ \epsilon^2\, k_y
\lbrace
\left( B_x-B_5 \sin{\phi}  \right)\, k_x + B_5 \cos {\phi} \, k_y
\rbrace \right]
\nn & = 
\frac{e \,  v_0\, v_z \, \Delta}{2 \, \epsilon^4} 
\left[
2 \, v_0^2 \, \kappa \cos{\gamma}
\cos{\phi} 
\lbrace   B_5 \cos^2 ({\phi}) - \left( B_x-B_5 \sin{\phi}  \right)  \sin{\phi} \rbrace
 - \frac{ \epsilon^2 \sin{\phi}
\lbrace \left( B_x-B_5\sin{\phi}  \right) \cos{\phi}  + B_5 \cos {\phi}\,  \sin {\phi} \rbrace }
{k_0 + \kappa \cos \gamma}  
\right],
 \nn  v^{\rm (m)}_y & = 
 \frac{ - e \, v_0\, v_z \, \Delta}{2\, \epsilon^4\, k_\perp^3} 
\left[ 2\, v_0^2\, k_y \left(k_\perp^2 - k_0\,k_\perp\right)\, 
\lbrace \left( B_x-B_5 \sin{\phi}  \right)\, k_y - B_5 \cos {\phi} \, k_x \rbrace - \epsilon^2\, k_x
 \lbrace \left( B_x-B_5 \sin{\phi}  \right)\, k_x + B_5 \cos {\phi} \, k_y \rbrace 
\right]\nn
& = 
\frac{ e \, v_0\, v_z \, \Delta}{2 \, \epsilon^4} 
\left[ 2 \,v_0^2 \, \kappa \cos {\gamma }
\sin {\phi}
\lbrace B_5 \cos^2 ({\phi}) - \left( B_x-B_5 \sin{\phi}  \right) \sin {\phi} \rbrace 
+ \frac{ 
\epsilon^2 \cos{\phi}
\lbrace B_5 \cos {\phi} \,\sin{\phi} +\left( B_x-B_5 \sin{\phi}  \right) \cos {\phi} \rbrace }
{k_0 + \kappa \cos {\gamma}}  
\right],
 \nn v^{\rm (m)}_z  & = 
\frac{ e \, v_0\, v_z^3 \, \Delta \,k_z }
{\epsilon^4\, k_\perp} 
\left[ B_5 \cos {\phi} \,k_x - \left( B_x-B_5 \sin{\phi} \right)\,k_y \right]
=
\frac{e  \, v_0^2 \, v_z^2 \, \Delta \, \kappa \sin {\gamma} } {\epsilon^4} 
\left[
B_5 \cos^2 ({\phi}) - \left( B_x-B_5\sin{\phi}  \right)  \sin{\phi} \right ].
\end{align}

\section{Details for the current density originating from the action of the Lorentz-force operator}
\label{appLF}

For the conductivity arising from the action of the LF operator, as shown in Eq.~\eqref{lf_cond1}, the solution is obtained by taking the terms (each term being labelled in the summation upto a chosen value of $n$). Thereafter, we need to expand the expressions in a given term upto the desired power in $| \bs B|$ --- we sort and label the sub-parts as $\mathcal N_{n, p}$, when that sub-part contains $| \bs B|^p$~\cite{ips-rsw-ph, ips-spin1-ph, ips_tilted_dirac}. Here we will assume that the net magnetic field is $\bs B$. The inclusion of strain for GNR is straightforward --- we just need to replace $\bs B$ by $( \bs B + \bs B_5 ) $. For the purpose of this paper, we have retained terms up to $n=3$. In this appendix, for the sake of completeness, we explicitly show the forms of these 3 contributions in 3 separate subsections.

\subsection{$n=1$: Terms originating from linear action of the Lorentz-force operator} 
\label{appn1}

The $n=1$ term leads to the current density of
\begin{align} 
& \bs{J}^{\rm{lf}}_{s} = 
-\,e^3 \,  \tau^2 \int \frac{d^3 \bs{k}} {(2 \, \pi)^3} 
\, \left[ {\boldsymbol v}_s 
+   {\boldsymbol u}_s \right]
\, D_s
\, f_{0}^\prime (\mathcal{E}_s)
\left( t_1+ t_2   \right ), \nn
& t_1= D_s \,
\check{L}
\left [    \left
\lbrace {\boldsymbol v}_s 
+ {\boldsymbol u}_s \right 
\rbrace \cdot \bs{E}   \right ] ,\quad
t_2 =\left [   \left
\lbrace {\boldsymbol v}_s 
+ {\boldsymbol u}_s \right 
\rbrace \cdot \bs{E}   \right ]
\check{L} \,
D_s\,.
\end{align}
In the following, we will use the new variables,
\begin{align}
{\boldsymbol V}^s &= e \left( {\boldsymbol v}^{(0,s)} \cdot \boldsymbol{\Omega}_s \right){\bs B}, \qquad
{\boldsymbol U}^{\rm (m)} = e \left( {\boldsymbol v}^{\rm (m)} \cdot \boldsymbol{\Omega}_s \right){\bs B}\,.
\end{align}
Expanding terms upto $\order{ | \bs B|^3}$, we obtain
\begin{align}
 t_1 &=
 \lbrace 1 -e \left (\boldsymbol{\Omega }_s \cdot   \bs{B} \right)
 + e^2   \left (\boldsymbol{\Omega }_s \cdot   \bs{B}  \right)^2 \rbrace 
  ({\boldsymbol v}^{(0,s)} \cross \bs{B}) \cdot \nabla_{\bs{k}}
  \left( {\boldsymbol v}^{(0,s)} \cdot \bs{E} \right )
+ ({\boldsymbol v}^{(0,s)} \cross \bs{B})
 \cdot \nabla_{\bs{k}} 
 \left( {\boldsymbol  U}^{\rm (m)}   \cdot \bs{E} \right ) \nn
 & \qquad  +\lbrace 1 -e \left (\boldsymbol{\Omega }_s 
 \cdot   \bs{B}  \right) \rbrace 
  ({\boldsymbol v}^{(0,s)} \cross \bs{B}) \cdot 
  \nabla_{\bs{k}}
  \left[ \left( \boldsymbol {v^{\rm (m)}  + V}^s \right) \cdot \bs{E} \right]
+ ({\boldsymbol  v}^{\rm (m)}  \cross \bs{B}) \cdot 
\nabla_{\bs{k}}\, \left[ \left( {\boldsymbol  v}^{\rm (m)}
 + {\boldsymbol V}^s \right) \cdot \bs{E} \right]  \nn
 & \qquad 
 +\lbrace 1 -e \left (\boldsymbol{\Omega }_s \cdot   
 \bs{B}  \right) \rbrace 
  ({\boldsymbol  v}^{\rm (m)}  \cross \bs{B}) \cdot \nabla_{\bs{k}} 
   \left(  {\boldsymbol  v}^{(0,s)}  \cdot \bs{E} \right ),
\end{align}

and
\begin{align} 
t_2 &=
\left ( {\boldsymbol v}^{(0,s)} \cdot \bs{E} \right)
 ({\boldsymbol v}^{(0,s)} \cross \bs{B})  \cdot \nabla_{\bs{k}}
 \left[  -e \left (\boldsymbol{\Omega }_s \cdot   \bs{B}  \right)
 + e^2   \left (\boldsymbol{\Omega }_s \cdot   \bs{B}  \right)^2 \right]
 + \left ( {\boldsymbol v}^{(0,s)} \cdot \bs{E} \right)
 ({\boldsymbol  v}^{\rm (m)}  \cross \bs{B})  \cdot \nabla_{\bs{k}}
 \left[  -e \left (\boldsymbol{\Omega }_s \cdot   \bs{B}  \right) \right] \nn
 & \qquad 
 +  \left ( {\boldsymbol v}^{\rm (m)} + {\boldsymbol V}^s\right)\cdot \bs{E} 
 \left({\boldsymbol v}^{(0,s)} \cross \bs{B}\right)  \cdot \nabla_{\bs{k}}
 \left[  -e \left (\boldsymbol{\Omega }_s \cdot   \bs{B}  \right) \right] .
\end{align}

Let us express the current density as
\begin{align} 
\label{n1}
& \bs{J}^{\rm{lf}}_s = 
-e^3 \,  \tau^2 \int \frac{d^3 \boldsymbol{k}} {(2 \, \pi)^3}  
\sum_{p =1}^3 
\boldsymbol {\mathcal N}_{1,p} \,,
\end{align}
where $\boldsymbol {\mathcal N}_{1, p} $ has a $| \bs B|^p $-dependence.
These evaluate to following expressions:
\begin{enumerate}
\item Linear-in-$| \bs B|$:
\begin{align}
\label{n11}
\boldsymbol {\mathcal N}_{1,1} &=
   {\boldsymbol v}^{(0,s)}\,  f^\prime_0  (\varepsilon_s )
   \left({\boldsymbol v}^{(0,s)} \cross \bs{B}\right)  \cdot \nabla_{\boldsymbol{k}} 
   \left ( {\boldsymbol v}^{(0,s)} \cdot \bs{E} \right) .
\end{align}

\item Quadratic-in-$| \bs B|$:
\begin{align}
\label{n12}
\boldsymbol {\mathcal N}_{1,2}  &={\boldsymbol v}^{(0,s)} 
   \left ( {\boldsymbol v}^{(0,s)} \cdot \bs{E} \right) f^\prime_0  (\varepsilon_s )  
    ({\boldsymbol v}^{(0,s)} \cross \bs{B})  \cdot \nabla_{\boldsymbol{k}}  
  \left[
-e \left (\boldsymbol{\Omega }_s \cdot  \bs{B}  \right) \right] 
   + {\boldsymbol v}^{(0,s)} \,  f^\prime_0  \left(\varepsilon_s  \right)
     \left({\boldsymbol v}^{(0,s)} \cross \bs{B}\right)  \cdot \nabla_{\boldsymbol{k}} 
     \left[  \left(
     {\boldsymbol  v}^{\rm (m)}   + {\boldsymbol V}^s \right) \cdot \bs{E} \right] 
   \nn
   &\quad
   +\left[ \left \lbrace -2 \, e \left( \boldsymbol{\Omega }_s \cdot  \bs{B}   \right) {\boldsymbol v}^{(0,s)}   
   + \left(  {\boldsymbol  v}^{\rm (m)} +  {\boldsymbol V}^s \right) \right \rbrace 
   f^\prime_0  (\varepsilon_s )
   - \left( \boldsymbol {m}_s \cdot \bs{B}   \right) {\boldsymbol v}^{(0,s)} 
   \,   f^{\prime \prime}_0  \left( \varepsilon_s  \right) 
\right]
\left( {\boldsymbol v}^{(0,s)} \cross \bs{B} \right)  \cdot \nabla_{\boldsymbol{k}} 
\left ({\boldsymbol v}^{(0,s)} \cdot  \bs{E}   \right ) \nn
   & \quad +{\boldsymbol v}^{(0,s)} \, f^{ \prime}_0  (\varepsilon_s ) 
   \left( {\boldsymbol  v}^{\rm (m)}  \cross \bs{B} \right)  \cdot \nabla_{\boldsymbol{k}} 
    \left( {\boldsymbol v}^{(0,s)} \cdot  \bs{E}   \right ).
 \end{align} 

\item Cubic-in-$| \bs B|$:
\begin{align}
\label{n13}
\boldsymbol {\mathcal N}_{1,3}  &=
\Big[
\left \lbrace -e \left(  \boldsymbol{\Omega}_s 
\cdot  \bs{B}\right)  {\boldsymbol v}^{(0,s)} 
\left( {\boldsymbol v}^{(0,s)} \cdot  \bs{E}  \right)
+ {\boldsymbol  v}^{(0,s)}  \left( {\boldsymbol  v}^{\rm (m)}  +{\boldsymbol V}^s   \right)
 \cdot  \bs{E} 
  + \left( {\boldsymbol v}^{(0,s)} \cdot  \bs{E}  \right)
     \left( {\boldsymbol  v}^{\rm (m)}  +{\boldsymbol V}^s   \right)
  \right \rbrace
      f^{ \prime}_0  (\varepsilon_s )
\nn & \qquad
 - \left( \boldsymbol {m}_s \cdot \bs{B}   \right) {\boldsymbol v}^{(0,s)}
       \left( {\boldsymbol v}^{(0,s)} \cdot  \bs{E}  \right)
  f^{\prime \prime}_0  (\varepsilon_ s) \Big]       \times
 \left[   \left({\boldsymbol v}^{(0,s)} \cross \bs{B}\right) 
    \cdot \nabla_{\boldsymbol{k}}
\left \lbrace -e \left( \boldsymbol{\Omega }_s 
\cdot  \bs{B}  \right)\right \rbrace \right ] \nn
 & \quad
 + {\boldsymbol v}^{(0,s)}  \left( {\boldsymbol v}^{(0,s)} \cdot  \bs{E}  \right) 
 f^{ \prime}_0  (\varepsilon_s )
     \left[ \left({\boldsymbol v}^{(0,s)} \cross \bs{B}\right) 
    \cdot \nabla_{\boldsymbol{k}}
    \left[ e^2 \left( \boldsymbol{\Omega }_s \cdot  \bs{B}    
  \right)^2 \right]
    + \left({\boldsymbol  v}^{\rm (m)}  \cross \bs{B}\right)
     \cdot \nabla_{\boldsymbol{k}}
    \left \lbrace -e \left( \boldsymbol{\Omega }_s \cdot  \bs{B} 
  \right) \right \rbrace  \right] 
 \nn & \quad  +
 \left({\boldsymbol v}^{(0,s)} \cross \bs{B}\right)  \cdot
   \nabla_{\boldsymbol{k}} \left ( {\boldsymbol v}^{(0,s)} \cdot \bs{E} \right)
     \Big[ \left \lbrace 3\, e^2\left( \boldsymbol{\Omega }_s 
  \cdot  \bs{B}    \right)^2  {\boldsymbol v}^{(0,s)}
     -2\, e \left( \boldsymbol{\Omega }_s \cdot  \bs{B} 
     \right)  \left({\boldsymbol  v}^{\rm (m)} +{\boldsymbol V}^s   \right)
     + {\boldsymbol  U}^{\rm (m)} \right \rbrace
      f^{\prime}_0  (\varepsilon_s )
 \nn & \hspace{ 1 cm }      
 + \Big\{ 2\, e \left( \boldsymbol{\Omega }_s \cdot  \bs{B}  \right)  {\boldsymbol v}^{(0,s)} 
- \left({\boldsymbol  v}^{\rm (m)} +{\boldsymbol V}^s   \right)
       \Big\}   \left( \boldsymbol{m}_s \cdot  \bs{B} \right)
        f^{ \prime \prime}_0  (\varepsilon_s )
     + \frac{ \left( \boldsymbol{m}_s \cdot  \bs{B} \right)^2}{2}
\, {\boldsymbol v}^{(0,s)}\, 
         f^{ \prime \prime \prime}_0  (\varepsilon_s )
  \Big] 
\nn & \quad +  
 \left[ \left \lbrace-2\, e \left( \boldsymbol{\Omega }_s \cdot  \bs{B}    \right) 
      {\boldsymbol v}^{(0,s)} 
      + \left({\boldsymbol  v}^{\rm (m)}  + {\boldsymbol V}^s    \right) \right \rbrace  
      f^{ \prime }_0  (\varepsilon_s )
      -\left( \boldsymbol{m}_s \cdot  \bs{B} \right) {\boldsymbol v}^{(0,s)} 
 \, f^{ \prime \prime}_0  (\varepsilon_s )
      \right]  \nn
& \qquad \times
\Big[ ({\boldsymbol v}^{(0,s)} \cross \bs{B})  
   \cdot \nabla_{\boldsymbol{k}} 
     \left \lbrace  
 \left ({\boldsymbol  v}^{\rm (m)}   + {\boldsymbol V}^s \right) 
 \cdot \bs{E}  \right \rbrace  
 +  ({\boldsymbol  v}^{\rm (m)}  \cross \bs{B})  \cdot \nabla_{\boldsymbol{k}} 
\left (  {\boldsymbol v}^{(0,s)} \cdot \bs{E} \right )
    \Big]  
\nn  & \quad
+ {\boldsymbol v}^{(0,s)} \,f^{ \prime}_0  (\varepsilon_s ) 
    \left[ ({\boldsymbol v}^{(0,s)} \cross \bs{B})
     \cdot \nabla_{\boldsymbol{k}}
\left (  {\boldsymbol  U}^{\rm (m)}  \cdot \bs{E} \right )
+  \left({\boldsymbol  v}^{\rm (m)}  \cross \bs{B}\right)
     \cdot \nabla_{\boldsymbol{k}} 
 \left \lbrace 
 \left  ({\boldsymbol  v}^{\rm (m)}   
 + {\boldsymbol V}^s \right) \cdot 
 \bs{E} \right \rbrace  \right].
\end{align}
\end{enumerate}

\subsection{$n=2$: Terms originating from quadratic action of the Lorentz-force operator} 
\label{appn2}

The $n=2$ term leads to the current density of
\begin{align} 
\bs{J}^{\rm{lf}}_s = 
-e^4 \,  \tau^3 \int \frac{d^3 \bs{k}} {(2 \, \pi)^3} 
 \lbrace {\boldsymbol v}_s 
+   {\boldsymbol u}_s \rbrace
 \left( D_s \right)^2
 f_{0}^\prime (\mathcal{E}_s)\,
\check{L}^2 \left [  D_s \left
\lbrace {\boldsymbol v}_s 
+{\boldsymbol u}_s  \right 
\rbrace \cdot \bs{E}   \right ] .
\end{align}
Due to the presence of $\check{L}^2$, there is no linear-in-$| \bs B|$ term here.

Let us express the current density as
 \begin{align} 
 \label{n2}
& \bs{J}^{\rm{lf}}_s = 
-\,e^4 \,  \tau^3 \int \frac{d^3 \bs{k}} {(2 \, \pi)^3}  
\sum_{ p =2}^3 \boldsymbol {\mathcal N}_{2,p} \,,
\end{align}
where $\boldsymbol {\mathcal N}_{2, p} $ has a $| \bs B|^p $-dependence.
These evaluate to following expressions:
\begin{enumerate}

\item Quadratic-in-$| \bs B|$:
\begin{align}
\label{n22}
\boldsymbol {\mathcal N}_{2, 2} &=
   {\boldsymbol v}^{(0,s)}\, f^\prime_0  (\varepsilon_s ) \left({\boldsymbol v}^{(0,s)} \cross \bs{B}\right) 
   \cdot \nabla_{\boldsymbol{k}}
   \left[
   \left({\boldsymbol v}^{(0,s)} \cross \bs{B}\right)  \cdot \nabla_{\boldsymbol{k}} 
   \left ( {\boldsymbol v}^{(0,s)} \cdot \bs{E} \right)\right].
\end{align}

\item Cubic-in-$| \bs B|$:
\begin{align}
\label{n23}
\boldsymbol {\mathcal N}_{2, 3} &=
     {\boldsymbol v}^{(0,s)}\, f^\prime_0  (\varepsilon_s )\,\left( {\boldsymbol v}^{(0,s)} \cross  \boldsymbol{B}  \right)
     \cdot
\nabla_{\boldsymbol{k}} 
 \Big[  -e \left(  \boldsymbol{\Omega }_s \cdot  \bs{B}\right) 
     \left( {\boldsymbol v}^{(0,s)} \cross  \bs{B}  \right)
     \cdot \nabla_{\boldsymbol{k}} \left ( {\boldsymbol v}^{(0,s)} \cdot \bs{E} \right)
 \nn & \hspace{ 5.25 cm}
 +  \left( {\boldsymbol v}^{(0,s)} \cross  \bs{B}  \right) 
\cdot \nabla_{\boldsymbol{k}} \left\lbrace \left ( {\boldsymbol v}^{\rm (m)} 
  +  {\boldsymbol V}^s \right)  \cdot \bs{E} \right \rbrace
+ \left( {\boldsymbol  v}^{\rm (m)}  \cross  \bs{B}  \right) \cdot 
      \nabla_{\boldsymbol{k}} \left ( {\boldsymbol v}^{(0,s)} \cdot \bs{E} \right) 
\Big]
\nn & \quad
+  {\boldsymbol v}^{(0,s)}\, f^\prime_0  (\varepsilon_s )
       \left( {\boldsymbol  v}^{\rm (m)}  \cross  \bs{B}  \right) \cdot 
      \nabla_{\boldsymbol{k}} \left[ \left( {\boldsymbol v}^{(0,s)} \cross  \bs{B}  \right) 
   \cdot \nabla_{\boldsymbol{k}} 
   \left( {\boldsymbol v}^{(0,s)} \cdot \bs{E} \right)  \right] 
\nn & \quad
 + {\boldsymbol v}^{(0,s)}\, f^\prime_0  (\varepsilon_s )
       \left( {\boldsymbol v}^{(0,s)} \cross  \bs{B}  \right) \cdot 
      \nabla_{\boldsymbol{k}} \left[ \left( {\boldsymbol v}^{(0,s)} \cdot \bs{E}  \right) 
      \left( {\boldsymbol v}^{(0,s)} \cross  \bs{B}  \right) \cdot 
      \nabla_{\boldsymbol{k}} \left \lbrace - e \left (  \boldsymbol{\Omega }_s \cdot 
      \bs{B} \right) \right \rbrace \right]\nn
& \quad
+\Big[ \left \lbrace -2\, e\, {\boldsymbol v}^{(0,s)} \left (  \boldsymbol{\Omega }_s 
\cdot \bs{B} \right) + 
{\boldsymbol  v}^{\rm (m)} 
+ {\boldsymbol V}^s \right \rbrace 
f^\prime_0  (\varepsilon_s )
-\left( \boldsymbol {m}_s \cdot \bs{B}   \right) {\boldsymbol v}^{(0,s)} \,  
 f^{\prime \prime}_0  (\varepsilon_s )\Big]\nn
 & \qquad \times
     \left({\boldsymbol v}^{(0,s)} \cross \bs{B}\right)
\cdot  
\left [\nabla_{\boldsymbol{k}}
  \left \lbrace  
\left( {\boldsymbol v}^{(0,s)} \cross  \bs{B}  \right) \cdot 
 \nabla_{\boldsymbol{k}} 
 \left (  {\boldsymbol v}^{(0,s)} \cdot \bs{E} \right) 
  \right \rbrace \right ].
\end{align}

\subsection{$n=3$: Terms originating from threefold action of the Lorentz-force operator} 
\label{appn3}
\end{enumerate}
The $n=3 $ term leads to the current density of
\begin{align} 
\bs{J}^{\rm{lf}}_s = 
-e^5 \,  \tau^4 \int \frac{d^3 \bs{k}} {(2 \, \pi)^3} 
\lbrace {\boldsymbol v}_s 
+   {\boldsymbol u}_s \rbrace
 \left( D_s \right)^3
 f_{0}^\prime (\mathcal{E}_s)\, 
\check{L}^3
\left [    D_s \left
\lbrace {\boldsymbol v}_s 
+{\boldsymbol u}_s  \right 
\rbrace \cdot \bs{E}   \right ] .
\end{align}

Due to the presence of $\check{L}^3 $, only a cubic-in-$| \bs B|$ needs to be extracted here. 
We can express the current density as
\begin{align} 
& \bs{J}^{\rm{lf}}_s = -\, e^5 \,  \tau^4
 \int \frac{d^3 \boldsymbol{k}} {(2 \, \pi)^3} \, 
 \boldsymbol {\mathcal N}_{3,3} \,,
\nn \text{where }
 \boldsymbol {\mathcal N}_{3, 3}  &=
   {\boldsymbol v}^{(0,s)}\, f^\prime_0  (\varepsilon_s ) 
  \left({\boldsymbol v}^{(0,s)} \cross \boldsymbol{B}\right) 
   \cdot \nabla_{\boldsymbol{k}} \left[
   \left({\boldsymbol v}^{(0,s)} \cross \boldsymbol{B}\right)  \cdot \nabla_{\boldsymbol{k}}
  \left \lbrace \left({\boldsymbol v}^{(0,s)} \cross \boldsymbol{B}\right)  
   \cdot \nabla_{\boldsymbol{k}}
   \left ( {\boldsymbol v}^{(0,s)} \cdot \boldsymbol{E} \right) \right \rbrace \right].
   \label{n33}
\end{align}

\section{Detailed results for the LF-induced parts}
\label{applfresults}

In this appendix, we demonstrate what the 3 terms, shown in Appendix~\ref{appLF}, evaluate to, when we consider the 3 set-ups for the GNR.

\subsection{Set-up I}
\label{appset1}

The individual expressions shown here feed into the final expressions summarised in Sec.~\ref{secset1}.
\begin{enumerate}
\item $n = 1$ --- This part comprises 3 sub-parts, which give us nonzero values for the out-of-plane conductivity.
$\bs{\mathcal{ N }}_{1,1}$ contributes to
\begin{align}
\sigma^{\text{lf,(h)}}_{zx} & = \frac{ \tau^2 \, e^3 \, v_0 \,v_z\,k_0\,B_y}
{8 \, \pi \, \mu^2}  
\left(\Delta^2-\mu^2  \right). 
\end{align}
$\bs{\mathcal{ N }}_{1,2}$  contributes to
\begin{align}
\sigma^{\text{lf,(bc)}}_{zx} & =
-\frac{\tau^2 \, e^{4} \, k_0 \, v_0^2 \, v_z^2 \, 
\Delta \left( \Delta^2 - \mu^2 \right)
B_5 \, B_y }
{4 \, \pi \, \mu^{5}}\,, \quad
\sigma^{\text{lf,(m)}}_{zx}  =
\frac{\tau^2 \, e^{4} \, v_0 \, v_z^2  \, \Delta \,B_5 \, B_y} 
{4 \, \pi \, \mu^{5}}
\left[k_0 \, v_0 \left( \Delta^2 - 3 \,\mu^2 \right)
+ \mu^2 \, \zeta \right]. 
\end{align}
Note that this one is absent for zero strain.
$\bs{\mathcal{ N }}_{1,3}$  contributes to
\begin{align}
\sigma^{\text{lf,(bc)}}_{zx} & =
\frac{9 \, \tau^2 \, e^{5} \, v_0^3 \, v_z^3 \, k_0 \,\Delta^2\, B_y}
{128 \, \pi \, \mu^{8}}\left[
\left(  B_x^2 + B_y^2 + 4\, B_5^2  \right)
\left(\Delta^2- \mu^2 \right)\right], \nn
\sigma^{\text{lf,(m)}}_{zx} & = \frac{\tau^2 \, e^5 \, v_0^2 \, v_z^3 \, \Delta^2\, B_y}
{128\, \pi \, \mu^8 \, \zeta}
\Big[
4\, B_5^2
\left\{
-27\, k_0\, v_0\, \Delta^2\, \zeta
- 4 \left( \Delta^2 + k_0\, v_0 \left( k_0\, v_0 - 7\, \zeta \right) \right) \mu^2
\right\}
\nn
& \hspace{3cm}
+ 3 \left( B_x^2 + B_y^2 \right) 
\left\{
-9\, k_0\, v_0\, \Delta^2\, \zeta
+ 2 \left( \Delta^2 + k_0\, v_0 \left( k_0\, v_0 + 3\, \zeta \right) \right) \mu^2
- 4\, \mu^4
\right\}
\Big], \nn
\sigma^{\text{lf, (conc)} }_{zx} & =  \frac{\tau^2 \, e^5 \, v_0^2 \, v_z^3 \, \Delta^2\, B_y}
{64\, \pi \, \mu^8}
\left[
4\, B_5^2 
\left\{
3\, k_0\, v_0\, \Delta^2
+ \left( 5\, k_0\, v_0 - 2\, \zeta \right) \mu^2\right\}
+ 3 \left( B_x^2 + B_y^2 \right)\left\{- \zeta\, \mu^2
+ k_0\, v_0 \left(\Delta^2 + 2\, \mu^2\right)\right\}\right]. \nn
\end{align}

\item $n = 2$ --- We find that nonzero contributions appear in the in-plane response, arising from 2 sub-parts.
$\bs{\mathcal{ N }}_{2,2}$ contributes to
\begin{align}
\sigma^{\text{lf,(bc)}}_{xx} & =  \sigma^{\text{lf,(m)}}_{xx} =  \sigma^{\text{lf, (conc)} }_{xx}  = 0\,,
\quad \sigma^{\text{lf,(h)}}_{xx}  =
\frac{\tau^3 \, e^{4} \,  v_z\, v_0^3 \, k_0  }
{16 \, \pi \, \mu^3}
\left[
2\, B_5^2 \left( \Delta^2-\mu^2  \right)
+ \left( B_x^2 + 3\, B_y^2 \right)
k_0 \, v_0 \left( \zeta - k_0 \, v_0  \right)
\right] \, , \nn
\sigma^{\text{lf,(bc)}}_{yx} & =  \sigma^{\text{lf,(m)}}_{yx} =  \sigma^{\text{lf, (conc)} }_{yx}  = 0\,,
\quad\sigma^{\text{lf,(h)}}_{yx} = \frac{\tau^3 \, e^4 \,v_z \, v_0^4\,k_0^2\,  B_x \, B_y  }
{ 8 \, \pi \, \mu^3}
  \left ( k_0\,v_0 - \zeta \right ).  
\end{align}
$\bs{\mathcal{ N }}_{2,3}$ contributes to
\begin{align}
\sigma^{\text{lf,(h)}}_{xx}  &= \sigma^{\text{lf, (conc)} }_{xx}  = 0, \nn
\sigma^{\text{lf,(bc)}}_{xx} &= - \, \frac{3\, \tau^3 \, e^5 \, v_z^2 \, v_0^4 \, k_0 \, \Delta \, B_5}
{64\, \pi \, \mu^6}
\left[4\, B_5^2 \left(\zeta^2- k_0^2 \, v_0^2 \right)
+ \left( B_x^2 + 3\, B_y^2 \right) \left(
\zeta^2 - 5\, k_0^2 \, v_0^2
+ 4\, k_0 \, v_0 \, \zeta \right)\right] ,\nn
\sigma^{\text{lf,(m)}}_{xx} & = \frac{\tau^3 \, e^5 \, v_z^2 \, v_0^3 \, \Delta \, B_5} {64\, \pi \, \zeta \, \mu^6}
\Big[
 B_x^2 \left\{12\, k_0^2 \, v_0^2 \, \zeta^2 + \left( 26\, k_0^2 \, v_0^2 + 12\, \Delta^2 \right) \mu^2- 12\, \mu^4
+ k_0 \, v_0 \, \zeta \left(
3\, \Delta^2 -12\, k_0^2 \, v_0^2- 35\, \mu^2\right)\right\}\nn
& \hspace{3 cm} 
+ B_y^2 \left\{ 36\, k_0^2 \, v_0^2 \, \zeta^2+ \left( 46\, k_0^2 \, v_0^2 + 36\, \Delta^2 \right) \mu^2- 36\, \mu^4
+ k_0 \, v_0 \, \zeta \left(
9\, \Delta^2-36\, k_0^2 \, v_0^2  - 73\, \mu^2\right)\right\} \nn 
& \hspace{3 cm} 
+4\, B_5^2 \, k_0 \, v_0 \, \zeta \left(3\, \Delta^2 - 5\, \mu^2\right)\Big]\,,
\end{align}
\begin{align}
\sigma^{\text{lf,(h)}}_{yx}  &= \sigma^{({\rm conc})}_{yx}  = 0, \quad%
\sigma^{\text{lf,(bc)}}_{yx}  =\frac{3\, \tau^3 \, e^5 \, v_z^2 \, v_0^4 \, k_0 \, \Delta \, B_5 \, B_x \, B_y}
{32\, \pi \, \mu^6}
\left(\zeta - k_0 \, v_0 \right)
\left( 5\, k_0 \, v_0 + \zeta \right),\nn
\sigma^{\text{lf,(m)}}_{yx} & = 
\frac{\tau^3 \, e^5 \,v_z^2 \,v_0^3 \,  \Delta \, B_5 \, B_x \, B_y  }  
{32 \pi \, \zeta \left(k_0^2 \,v_0^2 + \Delta^2 - \zeta^2\right)^3}
\Big[
  12\,\zeta^4 
  - 19 \, k_0\, v_0 \, \zeta^3 
  - \left(26 \,k_0^2\, v_0^2 + 12\,\Delta^2\right)\zeta^2 
  + k_0\, v_0 \left(31 \,k_0^2 \,v_0^2 + 16\,\Delta^2\right)\zeta \nn
  & \hspace{4.5 cm} 
+ 2 \,k_0^2 \,v_0^2 \left(k_0^2\, v_0^2 + \Delta^2\right) 
\Big].
\end{align}
These are nonzero only for nonzero strain.

\item $n = 3$ --- The sole nonzero conductivity component is the $zx$ component, sourced from terms arising from $\bs{\mathcal{ N }}_{3,3}$:
\begin{align}
\sigma^{\text{lf,(h)}}_{zx} &=\frac{
3\,\tau^4 \, e^5\, v_z^3 \, v_0^3\, k_0\, B_y 
} {16 \,\pi\, \mu^4 }
\left[
\left( B_x^2 + B_y^2 + 16\, B_5^2 \right)
k_0 \, v_0  \left(k_0 \,v_0 - \zeta\right)
+ 6\, B_5^2 \,(\Delta^2 - \mu^2)
\right].
\end{align}
Saliently, they receive no contribution from the BC or OMM.

\end{enumerate}

\subsection{Set-up II}
\label{appset2}

The individual expressions shown here feed into the final expressions summarised in Sec.~\ref{secset2}.
\begin{enumerate}
\item $n = 1 $ --- Only the out-of-plane conductivity is nonzero, sourced by 2 sub-parts.
$\bs{\mathcal{ N }}_{1,2}$ contributes to
\begin{align}
\sigma^{\text{lf,(h)}}_{yx}  &=   \sigma^{\text{lf,(bc)}}_{yx}  =\sigma^{\text{lf, (conc)} }_{yx}  = 0, \quad
\sigma^{\text{lf,(m)}}_{yx}  = - \frac{
e^4 \,  k_0 \, v_0^4 \, \Delta\,B_5 \, B_z \left(\zeta - k_0\, v_0\right)
\tau^2}{8 \, \pi \, \mu^3 \, \zeta}\,.
\end{align}
The last one is zero for zero strain.
$\bs{\mathcal{ N }}_{1,3}$ contributes to
\begin{align}
\sigma^{\text{lf,(h)}}_{yx}  &=  \sigma^{\text{lf,(bc)}}_{yx}  = 0, \quad
\sigma^{\text{lf},({\rm conc})}_{yx}  = - \frac{e^5 \,  k_0 \, v_z\, v_0^5 \, \Delta^2 \,B_5^2 \, B_z 
\left(\zeta - k_0\, v_0
\right)\tau^2}{16 \, \pi \, \mu^6 \, \zeta},\nn
\sigma^{\text{lf,(m)}}_{yx}  &  = \frac{
\tau^2\, e^5\, v_z \, v_0^4  \, \Delta^2B_z }{32 \, \pi  \, \mu^6\, \zeta^3}
\Big[
2 \, B_5^2 \, k_0 \, v_0
\left\{\zeta^2\left(- k_0 \, v_0 + \zeta\right)+ k_0 \, v_0 \, \mu^2\right\}
\nn & \hspace{3cm}
+ B_x^2\left\{3 \, \mu^2\left(\Delta^2 - \mu^2\right)+ 2 \, k_0 \, v_0\left(- \zeta^2
\left(- k_0 \, v_0 + \zeta\right)+ k_0 \, v_0 \, \mu^2\right)\right\}\Big]\,.
\end{align}

\item $n =2 $ --- Only the longitudinal components survive, arising from 2 sub-parts.
$\bs{\mathcal{ N }}_{2,2}$ contributes to
\begin{align} 
&\sigma^{\text{lf,(bc)}}_{xx} =  \sigma^{\text{lf, (conc)} }_{xx}  = \sigma^{\text{lf,(m)}}_{xx} = 0,\nn
& \sigma^{\text{lf,(h)}}_{xx} =\frac{\tau^3\,e^4 \,v_0^3\, k_0\left[2\, v_0^2\, B_z^2\left\{
-\Delta^2 + \mu^2+ 2 \,k_0^2 \,v_0^2\left(1 - \frac{k_0 \,v_0}{\zeta}\right)\right\}
+
v_z^2\left\{2\, B_5^2\, (\Delta^2 - \mu^2)
+ B_x^2\, k_0 \,v_0^2 \left( \frac{\zeta}{v_0} - k_0 \right)\right\}
\right]} {16 \,\pi \,v_z \,\mu^3}\,.
\end{align}
$\bs{\mathcal{ N }}_{2,3}$ contributes to
\begin{align}
\sigma^{\text{lf,(h)}}_{xx} & = \sigma^{\text{lf, (conc)} }_{xx}=0 \,,\nn
\sigma^{\text{lf,(bc)}}_{xx}& = 
\frac{\tau^3\, e^5\, v_0^4\,k_0\,\Delta \,(k_0\, v_0-\zeta) B_5} {64\,\pi\,\mu^6}
\left[\frac{16\,B_z^2\,v_0^2 \, (k_0\, v_0-\zeta) \, (2\, k_0  \,v_0+\zeta)}  {\zeta}
+3\,v_z^2\ \left\{4\,B_5^2 \, (k_0 \, v_0+\zeta)
  + B_x^2 \, (5\, k_0\, v_0+\zeta)   \right\}\right], \nn
 \sigma^{\text{lf,(m)}}_{xx} & = 
\frac{\tau^3 \,e^5\, v_0^3\, \Delta\, B_5} {64\,\pi\,\mu^6}
\Big[\frac{8\,B_z^2\,k_0\,  v_0^3 \,(k_0\,  v_0-\zeta)}{\zeta^3}
\left\{
-2\,  \zeta^2 \,(k_0\, v_0-\zeta) \,(2\,k_0\, v_0+\zeta)
+(k_0\, v_0- 2\, \zeta) \, (k_0 \,v_0 + 3\, \zeta)\,\mu^2 \right\}\nn
&\hspace{ 3 cm}
+ v_z^2\ \Big\{\frac{B_x^2}{\zeta} \left( 3 \,k_0\, v_0\,\zeta^3
+ 12\, \zeta^2 \left(k_0^2\, v_0^2+\mu^2\right)-\zeta \left(15\, k_0^3 \, v_0^3+32\, k_0 \, v_0\,\mu^2\right)
  +14\, k_0^2 \,v_0^2\,\mu^2 \right)\nn
&\hspace{3 cm}
-4\,B_5^2\,k_0 \, v_0 \left(3 \, k_0^2\, v_0^2-3\, \zeta^2+2\, \mu^2 \right)\Big\}\Big]\,,\nn
 \sigma^{\text{lf,(h)}}_{zx}  &= \sigma^{\text{lf},({\rm conc})}_{zx} =0 ,\quad
\sigma^{\text{lf,(bc)}}_{zx} = \frac{\tau^3 \, e^5 \, v_z^2  \, v_0^4 \, k_0\,
B_5 \, B_x \, B_z \,
\Delta \, \left( \zeta-k_0 \, v_0 \right)^2 }{8 \, \pi \, \mu^6}, \nn
\sigma^{\text{lf,(m)}}_{zx} & = 
\frac{
\tau^3 \, e^5  \, v_z^2\, v_0^3\,B_5 \, B_x \, B_z \Delta}{8 \, \pi \, \mu^6 \, \zeta}
\Big[ k_0 \, v_0\left\{
2 \,k_0 \, v_0\left(k_0^2 \, v_0^2 + \Delta^2\right)-\left(2\, k_0^2 \,v_0^2 + \Delta^2\right)
\zeta\right\} 
-
\left\{ 3 \,\Delta^2 + k_0 \, v_0 \left( 7 \,k_0 \, v_0 - 6 \,\zeta  \right)\right\}\mu^2 
+ 3\, \mu^4 \Big]\,. 
\end{align}
All these go to zero as $B_5$ goes to zero.
\item $n = 3$ --- Only the out-of-plane component of the response survives, sourced by  $\bs{\mathcal{ N }}_{3,3}$, which is given by
\begin{align}
\sigma^{\text{lf,(h)}}_{yx} &=-\frac{
 \tau^4\, e^5 \, v_0^5 \,  k_0\,B_z }{8 \, \pi \, v_z \, \mu^4\, \zeta^3 }
\Big[B_x^2 \, v_z^2 \, \zeta^2
\left\{-8\, k_0^2\, v_0^2
\left(k_0\, v_0 - \zeta\right)
+
\left(-5\, k_0 \,v_0 + \zeta\right)
\left(\Delta^2 - \mu^2\right)
\right\}\nn
&\hspace{3cm}+ 2\, B_z^2 \,v_0^2
\left\{4 \,k_0^4 \,v_0^4\left(
k_0\, v_0 - \zeta\right)
+ k_0^2 \,v_0^2\left(5 \,k_0 \,v_0 - 3\, \zeta\right)
\left(\Delta^2 - \mu^2\right)+ \zeta
\left(\Delta^2 - \mu^2\right)^2\right\}\nn
&\hspace{3cm}
+ 6\, B_5^2\, v_z^2\, \zeta^2\left\{4\, k_0^2 \,v_0^2\left(k_0 \,v_0 - \zeta\right)
+\left(\zeta -3\, k_0\, v_0 \right)
\left( \mu^2 - \Delta^2 \right)\right\}\Big]\,.
\end{align}
\end{enumerate}

\subsection{Set-up III}
\label{appset3}

The individual expressions shown here feed into the final expressions summarised in Sec.~\ref{secset3}.
\begin{enumerate}
\item $n = 1$ --- The surviving ones are the out-of-plane components, which arise from three parts.
$\bs{\mathcal{ N }}_{1,1}$ contributes to
\begin{align}
\sigma^{\text{lf,(h)}}_{yz} & = 
\frac{\tau^2 \, e^3 \, v_z\,v_0 \,k_0\,B_x}
{8 \, \pi \, \mu^2}  
\left(\Delta^2 - \mu^2  \right).
\end{align}
$\bs{\mathcal{ N }}_{1,2}$ contributes to
\begin{align}
\sigma^{\text{lf,(bc)}}_{yz}  =
-\frac{
 \tau^2\, e^4\,v_z^2\, v_0^2\, k_0\, \Delta\, 
B_5\, B_x \left( \Delta^2 - \mu^2 \right)}
{4\, \pi\, \mu^5} , \quad
\sigma^{\text{lf,(m)}}_{yz}   = 
\frac{
\tau^2\, e^4\,v_z^2\,  v_0^2\,k_0\, \Delta\,
B_5\, B_x  \left( \Delta^2 -2\, \mu^2 \right)}
{4\, \pi\, \mu^5}. 
\end{align}
$\bs{\mathcal{ N }}_{1,3}$ contributes to
\begin{align}
\sigma^{\text{lf,(bc)}}_{yz} & =
\frac{
9\,\tau^2 \, e^5 \, v_z^3 \, v_0^3  \, k_0\,\Delta^2 \, B_x 
\left( B_x^2 + 4 \, B_5^2 \right)}
{128 \, \pi \, \mu^8} 
  \left(\Delta^2 -\mu^2 \right),\nn
   \sigma^{\text{lf,(m)}}_{yz} &  = \frac{
3\, \tau^2 \, e^5\, v_z^3\, v_0^2 \,
\Delta^2  \, B_x  }{128 \, \pi \, \zeta \, \mu^8}
\left[
4 \, B_5^2 \, k_0 \, v_0 \, \zeta
\left( 8\, \mu^2 - 9\, \Delta^2 \right)
+ B_x^2\left\{-9\, k_0 \,v_0 \,\Delta^2\, \zeta
+ 2\, \zeta
\left(3\, k_0\, v_0 + \zeta\right)
\mu^2
- 2\, \mu^4
\right\}
\right],\nn
\sigma^{\text{lf, (conc)} }_{yz} &  =
\frac{
 \tau^2 \, e^5\, v_z \,v_0^2 \,\Delta^2 \,
B_x  }{64 \, \pi \, \mu^8 \,\zeta}
\Big[4\, k_0\, v_0\,\left\{
B_z^2\, v_0^2
\left(k_0\, v_0 - \zeta\right)\mu^2
+3\, B_5^2\, v_z^2 \,\zeta
\left(
\Delta^2 + \mu^2\right)\right\}
 \nn &\hspace{ 3.5 cm}
-3\, B_x^2 \,v_z^2 \,\zeta\left\{k_0^3\, v_0^3
+\zeta \,\mu^2-k_0\, v_0
\left(\zeta^2 + 3\, \mu^2\right)\right\}\Big].
\end{align}

\item $n =2 $ ---
The sole surviving part is the longitudinal component captured by 2 parts.
$\bs{\mathcal{ N }}_{2,2}$ contributes to
\begin{align}
\sigma^{\text{lf,(h)}}_{zz} = \frac{  \tau^3 \, e^4 \, v_0 \,v_z^3\,k_0 \left( B_x^2 + 2 \, B_5^2  \right)}
{8 \, \pi \, \mu^3}  
\left( \Delta^2-\mu^2 \right).
\end{align}
$\bs{\mathcal{ N }}_{2,3}$ contributes to
\begin{align}
\sigma^{\text{lf,(bc)}}_{xz} & = -\,
\frac{
\tau^3 \, e^5 \,v_z^2 \, v_0^4 \, k_0 \,  \Delta \,B_5 \, B_x \, B_z 
\left[ \Delta^2 - \mu^2
+ 2\, k_0 \,v_0 \left( k_0\, v_0 - \zeta \right)
\right]} {4 \, \pi \, \mu^6},\nn
\sigma^{\text{lf,(m)}}_{xz} &  =
-\,\frac{\tau^3 \, e^5 \, v_z^2 \, v_0^4 \,  k_0 \,\Delta \,B_5 \, B_x \, B_z  
\left[ k_0 \,v_0 
\left(2\, k_0^2 \,v_0^2+ 2\,\Delta^2-3\, \mu^2 \right)
- \zeta \left( 2\, k_0^2\, v_0^2 +  \Delta^2-2\, \mu^2 \right)
\right]}{4 \, \pi \, \mu^6\,\zeta}\,,
\end{align}
\begin{align}
\sigma^{\text{lf,(bc)}}_{zz} & =\frac{ \tau^3\, e^5 \, v_z^2 \, v_0\, \Delta \,B_5  
\left( -k_0 \,v_0 + \zeta \right)}{8 \, \pi \, \mu^6}
\Big[ B_z^2\, k_0\, v_0^3 \left( \zeta - k_0 \,v_0 \right)
\nn &  \hspace{3.8cm}
+ 3\, v_z^2 \left\{ - B_5^2 \, k_0 \,v_0 \left( k_0 \,v_0 + \zeta \right)
+ B_x^2 \left(
\zeta^2 -2 \,k_0^2 \,v_0^2 - 2 \,k_0 \,v_0 \zeta  \right) \right\}\Big]\,, \nn
\sigma^{\text{lf,(m)}}_{zz} &  =
\frac{\tau^3 \, e^5 \,v_z^2 \,v_0 \, \Delta \, B_5  }{8 \, \pi \, \mu^6}
\Big[B_z^2 \,k_0\, v_0^3
\left\{ \mu^2 -\Delta^2 + 2 \,k_0 \,v_0
\left( \zeta - k_0 \,v_0 \right) \right\}
\nn &\hspace{3 cm}
+ v_z^2\left\{B_5^2\, k_0 \,v_0\left( 3 \,\Delta^2 - 5\, \mu^2 \right)
+ 3\, B_x^2\left( \zeta\, \mu^2 -\zeta^3 +  k_0 \,v_0\left(k_0^2\, v_0^2+ 3 \,\Delta^2- 5 \,\mu^2\right)\right)
\right\}\Big]\,.
\end{align}

\item $n = 3 $ --- We find that an out-of-plane transverse component is the sole nonzero part, sourced by $\bs{\mathcal{ N }}_{3,3}$, which is
\begin{align}
\sigma^{\text{lf,(h)}}_{yz} &= 
\frac{
\tau^4 \, e^5 \,v_z \, v_0^3\,  k_0 \,B_x } {16 \, \pi \, \mu^4\, \zeta^2 }
\left[ 2\,  B_z^2\,  v_0^2 \, \zeta
\left( \zeta - k_0 \, v_0 \right)^2 \left( 2\,  k_0 \, v_0 + \zeta \right)
- 3\,  v_z^2 \, \zeta^2\left\{B_x^2\,  k_0\,  v_0\left( \zeta - k_0\,  v_0  \right)
+ 2 \, B_5^2\left( \Delta^2 - \mu^2 \right)\right\}\right].
\end{align}
\end{enumerate}

\bibliography{ref_nl}

\end{document}